\documentclass[twocolumn]{aastex631}
\usepackage{multirow}
\usepackage{soul}

\shorttitle{Optical polarization study towards Czernik3}
\shortauthors{Uppal et al.}
\graphicspath{{./}{figures/}}

\begin{document}

\title{Optical linear polarization study towards Czernik 3 open cluster at different spatial scales}

\author[0000-0001-5814-4558]{Namita Uppal}
\affiliation{Physical Research Laboratory, Ahmedabad, India}
\affiliation{Indian Institute of Technology, Gandhinagar, India}
\author[0000-0002-7721-3827]{Shashikiran Ganesh}

\affiliation{Physical Research Laboratory, Ahmedabad, India}

\author[0000-0002-8988-8434]{D. Bisht}
\affiliation{Key Laboratory for Research in Galaxies and Cosmology, University of Science and Technology of China, Chinese Academy of Sciences, Hefei, Anhui, China}

\begin{abstract}
We present the optical linear polarization observation of stars towards the core of the Czernik 3 cluster in the 
\textit{Sloan i}-band.  The data were obtained using the EMPOL instrument on the 1.2 m telescope at Mount Abu  Observatory.  We study the dust distribution towards this cluster by combining the results from our polarization observations with the data from Gaia EDR3, WISE, and the HI, $^{12}$CO surveys. In addition, we use the polarimetric data of previously studied clusters within 15$^\circ$ of Czernik 3 to understand the large scale dust distribution. The observational results of Czernik 3 show a large range in the degree of polarization, indicating that the dust is not uniformly distributed over the plane of the sky, even on a small scale. The distance to the Czernik 3 is constrained to $3.6\pm0.8$ kpc using the member stars in the core region identified from Gaia EDR3 astrometry. This makes it one of the most distant clusters observed for optical  polarization so far. The variation of observed degree of polarization and extinction towards this cluster direction suggest the presence of at least two dust layers along this line of sight at distances of $\sim1$ kpc and $\sim3.4$ kpc. There is an indication of the presence of dust in the centre of the cluster as seen from an increase in the degree of polarization and WISE W4 flux.  The large scale distribution of dust reveals the presence of a region of low dust content between the local arm and the Perseus arm. 

\end{abstract}

\keywords{(ISM:) dust, extinction- ISM: general- (Galaxy:) open clusters and associations: individual (Czernik 3)-methods: observational- techniques: polarimetric }


\section{Introduction} \label{sec:intro}

\label{sec:int}
Dust is an integral part of the interstellar medium (ISM) and is present everywhere in the sky.  Polarization and the extinction of star light are manifestations of the dust in the ISM. Starlight polarization by the interstellar dust is considered as a useful tool to determine the properties of the ISM, e.g., the geometry of the magnetic field at small scale \citep{goodman1990optical} as well as large scale \citep{mathewson1970polarization, axon1976catalogue, largescale, GPIPS}; grain properties: size, shape, chemical composition and grain efficiency \citep{mathis1986, kim1994size,kim1995size, clayton2003size,efficiency, NewSerkowski}. Besides this, polarization can also be used to evaluate the dust distribution along the line of sight \citep[e.g.,][]{Bk59}. The dust distribution in the Galaxy can be inferred from  extinction properties of the dust as well. Many attempts have been made to construct    3-dimensional extinction maps of the Milky Way galaxy \citep[][etc]{Lallement2019, Green2019}. However, these extinction maps are based on models and assumptions.  Hence these maps have an inherent bias.  On the other hand, polarization measurements give the direct observational signature of dust patches along the line of sight.

Open clusters are important constituents of the Galactic disk and their distribution has been studied to trace the spiral arm structure of the Galaxy \citep[][etc]{spiral1, spiral2}. They are good candidates to carry out the polarization observations as all the member stars have common properties in terms of their distance, age, and proper motion. Polarization data of different clusters will help to study the dust distribution in the Galactic disk particularly the spiral arms.
 Cluster polarization studies have been used to identify cluster members \citep{membership} and to understand the dust properties \citep[e.g.,][and references therein]{NGC654, NGC1893,  Bk59, NGC1502, Alessi1, NGC1817}. 
Linear polarization study of open clusters combined with the distance information from Gaia early data release 3 \citep[EDR3, ][]{EDR3} can help to trace the dust distribution. 

In this paper, we present a polarimetric study towards the core of the open cluster Czernik 3 [$\alpha_{J2000}: 01^h03^m06^s$ and $\delta_{J2000}: +62^\circ 47^\prime 00^{\prime\prime}$ \citep{Dias2002}].
The cluster is located in the 2$^{nd}$ Galactic quadrant ($\ell = 124^\circ.256, b=-0^\circ.058$).  Previous studies have provided a range of ages (e.g., 100 Myr: \citet{Dias2002}; 630 Myr: \citet{Kharchenko2016}; 115 Myr: \citet{Bisht}), photometric distances (1.4-1.75 kpc),  reddening E(B-V) (0.99-1.4 mag), and average extinction $A_{V} = 2.9$ mag \citep{Dias2002, Bisht, Kharchenko2016}. Recently, \citet{gaiaDR2OC} have derived basic parameters and membership of 1229 Galactic open clusters using parallax and proper motion information from Gaia DR2. They showed that this cluster has 48 members with membership probability $>$ 0.5, and about half of the member stars of this cluster lie within a radial distance of $0.82^\prime$.
They also estimated the most probable distance to the cluster to be 4.47 kpc. Czernik 3 cluster has an extended morphology \citep{CZ3sharma2020} with a radius of the core and the whole cluster $\sim 0.5-0.6^\prime$ and $1.2-5^\prime$, respectively \citep{Dias2002, Kharchenko2016, JoshiGC, Bisht, CZ3sharma2020}.  
\citet{CZ3sharma2020} have placed the cluster at an average distance of $3.5 \pm 0.9$ kpc. They have claimed that it is a disintegrating old cluster with an average age of $0.9^{+0.3}_{-0.1}$ Gyr. Thus, there is a large variation in the possible ages and distances to this cluster.  Astrometric data from Gaia EDR3 is expected to help better constrain the distance and membership.

The aim of our study is to understand the dust distribution  along the line of sight towards Czernik 3 at various spatial scales spanning a galactic longitude range of 110$^\circ$ to 140$^\circ$.  To this end, we complement our observations of Czernik 3 with other archival polarimetric observations.
We have redefined the cluster membership based on Gaia EDR3 data to improve the distance estimates.

This paper is organized as follows -  Section \ref{sec:2} describes our observations and the data reduction procedures followed.  It also includes details of the archival data used in this work. Results obtained from the polarization observations are presented in Section \ref{sec:3}. 
Section \ref{sec:4} consists of the discussion and is in four parts:  the first one discusses the cluster membership from Gaia EDR3 data.  The second and third subsections covers the polarization observations of Czernik 3 in the context of multi-wavelength information and the distribution of dust towards the cluster.  The fourth sub-section discusses the overall dust distribution by combining all available polarization data to various clusters and field stars towards a wider region centered on Czernik 3.
We conclude with a summary in Section \ref{sec:5}.
\section{Observations and archival data}\label{sec:2}
In this section, we describe our polarimetric observations and the data reduction procedures followed.  We also briefly discuss the archival data used in this study.
\subsection{Polarimetric observations and reductions}
Polarization observations of Czernik 3 were carried out on the dark nights of 2021, January 13 and February 7 using the EMPOL instrument \citep{EMPOL}. The instrument is mounted at the Cassegrain focus of the 1.2 m, f/13 telescope at the Mount Abu Observatory operated by the Physical Research Laboratory (PRL).  EMPOL is an EMCCD based optical imaging polarimeter, built in-house at PRL.  Polarization measurement is achieved using a rotating half wave plate as modulator and a wire grid polarizer as  analyzer. A fixed half-wave plate is also used just below the rotating half-wave plate to compensate for the wavelength dependence of the position angle of the half-wave plate.  This pair of identical half-wave plates are as per the superachromatic wave plate design provided by  \citet{panch}. Use of the superachromatic half-wave plates ensures that there is no wavelength dependence of  position angle. The rotating half waveplate completes one rotation in 48 steps of $7.5^\circ$ each. 

Andor iXon EMCCD detector used in EMPOL has $1K \times 1K$ pixels of 13 $\mu$m size with a plate scale of $0.18 ^{\prime\prime}$~pixel$^{-1}$ at the focal plane. We have used  $4\times 4$ on-chip binning to get a final plate scale of $0.72^{\prime\prime}$~pixel$^{-1}$ with a field of view of  $\sim3^{\prime} \times \sim3^{\prime}$.  On-chip binning and electron multiplicative gain (EMGAIN) of 20 is used in order to get enough counts for the faint stars. 

The observations were taken in \textit{Sloan  i} ($0.767\; \mu m$) filter with 0.5 sec exposures at each step of the half wave plate, covering the entire core region \citep[$0.5^\prime$-$0.6^\prime$;][]{kharchenko2013, CZ3sharma2020} in one pointing.  A series of exposures were taken to get a total of 40 sec effective exposure corresponding to each half waveplate step angle.  
Polarized standard HD25443 (see Table \ref{tab:standard} for observed values) and un-polarized standard HD12021 were also observed on each night along with the cluster field, in the same filter, to calibrate the polarization angle and to check the instrumental polarization respectively.  The instrumental polarization was found to be negligible (below 0.1\%) as seen by the measured degree of polarization, 0.16$\%$ $\pm$ 0.12$\%$ of the unpolarized standard star (HD12021).  The calibration/performance of the instrument for 100\% polarized light was tested using a glan prism in the light path prior to the rotating half-wave plate with a resulting polarization value of $\sim99$\%.

\begin{table*}[!ht]
    \caption{Observed and reference polarization values \citep{HD25443} of polarized standard star: HD25443 (Vmag = 6.78 mag).}\label{tab:standard}
    \begin{tabular}{|c|c|c|c|c|c|c|}
    \hline
    Observation epoch & $P_{obs}$ ($\%$)  & $\theta_{obs}$ ($^\circ$) & SNR\footnote{for a single position of half-wave plate} & $P_{ref}$ ($\%$) & $\theta_{ref}$ ($^\circ$) &  $\theta_{off}$ ($^\circ$)\\
    \hline
    Jan 13, 2021 & 4.16 $\pm$ 0.06 & 18.03 $\pm$ 0.42 & 160 & \multirow{2}{*}{4.33} & \multirow{2}{*}{134.21 $\pm$ 0.28} & 116.18 \\
    Feb 7, 2021 & 4.43 $\pm$ 0.19 & 22.07 $\pm$ 1.22 & 130 &  &  & 112.14\\
    \hline
    \end{tabular}
\end{table*}

The observed data were reduced and analysed using self scripted python routines. The basic data reduction tasks include - bias subtraction, flat-fielding followed by shifting and stacking of images to increase the signal to noise ratio.  The shifting and stacking was done in cycles of 48 frames to obtain a final stack of frames corresponding to the 48 step angles of the half-wave plate.  The image coordinates of stars in the field were obtained using \textit{SExtractor} software \citep{sextractor}.  Photometry on the extracted co-ordinates was carried out by aperture photometry method using photutils package \citep{photutils} of \textit{Astropy} \citep{astropyII}. A constant  optimal aperture i.e., 3 $\times$ FWHM, was used to carry out the photometry of the final 48 images.  In a crowded field, there is a finite probability for the aperture to overlap the close-by stars.  In order to minimize the effects of the overlap we have applied aperture corrections, i.e., chosen the photometric results from smaller apertures of 2 $\times$ FWHM scaled to 3 $\times$ FWHM using a scaling relation derived from the isolated stars.
The final photometric results from the 48 measurements (corresponding to the half-wave plate step angles) for each star are fitted according to eq. (\ref{1}):
\begin{equation}\label{1}
    I_j^\prime = \frac{1}{2} [I\pm Q\cos4\alpha_j \pm U\sin4\alpha_j]
\end{equation}
This expression is obtained by solving the Mueller matrices for the active optical elements in the instrument. Here, the I, Q, U are the best fit values of the Stokes parameters of the light incident on the instrument and $\alpha_j$ is the angle of the half-wave plate,  $0 \le j \le 47$.  $I_j^\prime$ represents the total intensity of the light recorded by the detector corresponding to $\alpha_j$ angle. The polarization fraction (P$_{obs}$) and polarization angle ($\theta$) were calculated from the Stokes parameters of the incoming light beam using eq. (\ref{2} \& \ref{3})
\begin{equation}\label{2}
P_{obs} = \frac{\sqrt{Q^2+U^2}}{I}
\end{equation}
\begin{equation}\label{3}
\theta = \frac{1}{2}\tan^{-1}\left(\frac{U}{Q}\right)
\end{equation}
  The errors in the degree of polarization ($\sigma_P$, eq. \ref{4}) and polarization angle ($\sigma_{\theta}$, eq. \ref{5}) were derived using fundamental error propagation methods:
\begin{equation}\label{4}
 \sigma_{P} = \frac{1}{I} \sqrt{\frac{Q^2\sigma_{Q}^2 + U^2\sigma_{U}^2}{Q^2+U^2} + \frac{Q^2+U^2}{I^2}\sigma_I^2}
\end{equation}
\begin{equation}\label{5}
 \sigma_{\theta} = \frac{1}{2(Q^2+U^2)} \sqrt{Q^2\sigma_{U}^2 + U^2\sigma_{Q}^2}
\end{equation}
We have also debiased the  polarization measurements using equation (\ref{eq:debias}) in order to compensate the biasing produced due to low signal-to-noise ratio (SNR) in some of the objects.
\begin{equation}\label{eq:debias}
    {P} = {\sqrt{P_{obs}^2 - \sigma_{P}^2}}
\end{equation}
All further analysis consider the debiased polarization values.
The observed position angles ($\theta$) were corrected to the  reference position angle using observations of the polarized standard star HD25443. The reference degree of polarization ($P_{ref}$) of HD25443 star in \textit{Sloan-i} band was not available in the literature. So, we calculated the value (provided in Table \ref{tab:standard}) from the Serkowski law of interstellar polarization by using $P_{max}$, $\lambda_{max}$, and K values for HD 25443 listed in \citet{HD25443}.
The astrometric calibration of the polarization images were carried out using the \textit{imwcs}\footnote{http://tdc-www.harvard.edu/wcstools/imwcs/}  program of \textit{World Coordinate System Tools} \citep[WCSTools,][]{WCSTOOL} software with the \textit{UCAC4} \citep{UCACcat} reference catalog. The astrometric error was found to be less than $0.5^{\prime\prime}$.  

We were able to derive the polarization measurements for 43 stars in the field from our observations.  One of these, star \#13, is a very bright foreground star and its glare has affected the observation of several fainter stars in its immediate vicinity (within 18$^{\prime\prime}$ radial distance). Therefore, we do not consider this star and its near neighbours in our analysis. We also do not consider the three stars (\#1, \#7, and \#42) having SNR $< 3$ for polarization.
\subsection{Archival Data}
We have used the astrometric and photometric data from Gaia EDR3 \citep{EDR3,GaiaEDR3Cat}. 
 We consider the distance (`rpgeo') derived from Gaia EDR3 by \citet{bailer2021} using Bayesian analysis method instead of inverse of parallax because around 50\% of these 42 stars have  fractional parallax error $ > 0.2 $. These distances have been used in our further analysis.
We have also used imaging and photometric data from Pan-STARRS \citep{panstarrs1}, 2MASS \citep{2MASS}, and WISE \citep{WISEW4} in order to compliment our polarimetric observations of Czernik 3 cluster.  The photometric and astrometric data were cross-matched with the stars observed by us, using CDS crossmatch service \citep{CDS} within 1$^{\prime\prime}$ search radius.   In addition, we have also used the spectral information of neutral hydrogen (HI) and $^{12}$CO from the HI4PI \citep{HI4PI} and $^{12}$CO \citep{12CO} surveys.

In order to study the large scale dust distribution, we have made use of polarimetric observations from the literature. \citet{Heiles2000} provide a list of polarization observations of approximately 9000 bright stars distributed in the plane of the Galaxy. 
From this compilation, we have used the stars within 15$^\circ$ region of Czernik 3. 
Five clusters with polarization measurements were also found within 15$^\circ$ of Czernik 3:  IC 1805,   NGC 654, Berkeley 59, NGC 457, and Alessi 1
(further details and references are in Table \ref{tab:clusters}). 

\begin{deluxetable*}{lclccccc}
\tablecaption{Details of the nearby open clusters around Czernik 3. The average angular radius is obtained from \citet{kharchenko2013} and average distance from \citet{gaiaDR2OC} \label{tab:clusters}}
\tablewidth{0pt}
\tablehead{
\colhead{Cluster} & \colhead{$\ell$} & \colhead{b} & \colhead{Angular distance} & \colhead{Reference} & \colhead{Angular radius} & \colhead{Average distance} \\ 
 & \colhead{$(^\circ)$} & \colhead{$(^\circ)$} & \colhead{from Cz3 ($^\circ$)} &  &
\colhead{($^\prime$)} & \colhead{($kpc$)}
}
\startdata
NGC 457 & 126.631 & -4.390 & 4.90 & \citet{NGC457} & 13.2 & 2.88\\
Berkeley 59 & 118.230 & 5.020 & 7.86 & \citet{Bk59} & 13.2 & 1.06 \\
NGC654 & 129.008& -0.359 & 4.80 & \citet{NGC654} & 9.6 & 2.92 \\
IC1805 & 134.733 & 0.945 &10.00 & \citet{IC1805} & 16.2 & 2.09 \\
Alessi1 & 123.255 & -13.330 & 13.00 & \citet{Alessi1} & 25.5 & 0.70\\
\enddata
\end{deluxetable*}
\section{Results of the polarization observations of Czernik 3}\label{sec:3}
The polarimetric results of the 43 stars observed towards Czernik 3 cluster in \textit{Sloan i}-band are listed in Table \ref{tab:obs}. Serial number (ID) and astrometric positions ($\alpha_{J2000}\; \&\;  \delta_{J2000}$) obtained from \textit{WCSTools} are given in column 1, 2, and 3 respectively. Columns 4-12 represent the degree of polarization (P$_{obs}$), debiased degree of polarization (P), polarization angle ($\theta$) (measured from north increasing towards east), normalized Stokes $q$, $u$ parameters, and their respective errors ($\epsilon_{P}, \epsilon_{\theta}, \epsilon_q, \epsilon_u$). The maximum degree of polarization obtained in \textit{Sloan i}-band is $5.89\% \pm 0.15\%$ with the  polarization angle of $75^\circ.7 \pm 0^\circ.7$. The weighted average polarization (P) and polarization angle ($\theta$) for  the 39 stars considered for the analysis, are $2.42\%$ and $80^\circ$ with a large dispersion $1.18 \%$ in P and $\sim16^\circ$ in $\theta.$

\begin{deluxetable*}{cccccccccccc}
\tablecaption{Polarization observations towards Czernik 3.  Details are in the text. ID with asterisk (*) marks the stars having polarization SNR $<$ 3. \label{tab:obs}}

\tablewidth{0pt}
\tablehead{
\colhead{ID} & \colhead{$\alpha_{J2000} (^\circ)$} & \colhead{$\delta_{J2000} (^\circ)$} & \colhead{$P_{obs} (\%)$} & \colhead{$\epsilon_{P} (\%)$} &\colhead{P ($\%$)} & \colhead{$\theta (^\circ)$} & \colhead{$\epsilon_{\theta} (^\circ)$} & \colhead{$q (\%)$} & \colhead{$\epsilon_{q}(\%)$} & \colhead{$u (\%)$} & \colhead{$\epsilon_{u}$}
}
\startdata
1$^*$ & 15.73366 & 62.78726 & 0.87 & 0.28 & 0.82 & 76.0 & 9.4 & -0.77 & 0.28 & 0.41 & 0.28\\
  2 & 15.73596 & 62.80278 & 1.23 & 0.34 & 1.18 & 128.2 & 7.8 & -0.29 & 0.34 & -1.2 & 0.34\\
  3 & 15.73702 & 62.80149 & 0.91 & 0.27 & 0.87 & 133.5 & 8.5 & -0.05 & 0.27 & -0.91 & 0.27\\
  4 & 15.73735 & 62.79727 & 1.22 & 0.24 & 1.20 & 84.3 & 5.6 & -1.2 & 0.24 & 0.24 & 0.24\\
  5 & 15.74000 & 62.79577 & 1.05 & 0.26 & 1.02 & 73.0 & 7.0 & -0.87 & 0.26 & 0.59 & 0.26\\
  6 & 15.74345 & 62.79857 & 0.91 & 0.27 & 0.87 & 97.2 & 8.5 & -0.88 & 0.27 & -0.23 & 0.27\\
  7$^*$ & 15.74398 & 62.78652 & 0.4 & 0.25 & 0.31 & 75.8 & 17.6 & -0.35 & 0.25 & 0.19 & 0.24\\
  8 & 15.74801 & 62.79844 & 1.31 & 0.26 & 1.28 &  90.8 & 5.6 & -1.31 & 0.26 & -0.03 & 0.26\\
  9 & 15.74874 & 62.79615 & 1.16 & 0.24 & 1.13 & 83.5 & 6.0 & -1.13 & 0.24 & 0.26 & 0.24\\
  10 & 15.75125 & 62.78201 & 2.42 & 0.2 & 2.41 & 67.1 & 2.4 & -1.69 & 0.2 & 1.74 & 0.2\\
  11 & 15.75418 & 62.78105 & 2.11 & 0.19 & 2.10 & 79.6 & 2.6 & -1.97 & 0.19 & 0.75 & 0.19\\
  12 & 15.75874 & 62.78557 & 2.01 & 0.21 & 2.00 & 74.0 & 2.9 & -1.7 & 0.21 & 1.07 & 0.21\\
  13 & 15.76230 & 62.79320 & 1.83 & 0.18 & 1.82 & 87.0 & 2.8 & -1.82 & 0.18 & 0.19 & 0.18\\
  14 & 15.76590 & 62.80123 & 1.04 & 0.19 & 1.02 & 94.5 & 5.3 & -1.03 & 0.19 & -0.16 & 0.19\\
  15 & 15.76660 & 62.78815 & 1.81 & 0.25 & 1.79 & 75.7 & 3.9 & -1.59 & 0.25 & 0.87 & 0.25\\
  16 & 15.76776 & 62.78504 & 3.29 & 0.15 & 3.29 & 76.4 & 1.3 & -2.93 & 0.15 & 1.5 & 0.15\\
  17 & 15.76883 & 62.77100 & 1.13 & 0.22 & 1.11 & 81.2 & 5.5 & -1.08 & 0.22 & 0.34 & 0.22\\
  18 & 15.76958 & 62.78195 & 1.72 & 0.25 & 1.70 & 78.0 & 4.1 & -1.58 & 0.25 & 0.7 & 0.25\\
  19 & 15.77562 & 62.78632 & 3.45 & 0.16 & 3.45 & 74.5 & 1.3 & -2.95 & 0.16 & 1.78 & 0.16\\
  20 & 15.77565 & 62.78210 & 4.25 & 0.15 & 4.25 & 75.9 & 1.0 & -3.74 & 0.15 & 2.01 & 0.15\\
  21 & 15.77981 & 62.78914 & 4.01 & 0.17 & 4.01 & 75.0 & 1.2 & -3.47 & 0.17 & 2.0 & 0.16\\
  22 & 15.78057 & 62.78682 & 1.83 & 0.27 & 1.81 & 75.6 & 4.2 & -1.6 & 0.27 & 0.88 & 0.27\\
  23 & 15.78078 & 62.80348 & 1.3 & 0.23 & 1.28 & 129.0 & 4.9 & -0.27 & 0.22 & -1.27 & 0.23\\
  24 & 15.78158 & 62.77566 & 1.02 & 0.27 & 0.98 & 79.7 & 7.6 & -0.96 & 0.27 & 0.36 & 0.27\\
  25 & 15.78315 & 62.78855 & 2.41 & 0.22 & 2.40 & 76.8 & 2.6 & -2.16 & 0.22 & 1.07 & 0.22\\
  26 & 15.78333 & 62.78564 & 1.93 & 0.2 & 1.92 & 73.8 & 2.9 & -1.63 & 0.2 & 1.03 & 0.2\\
  27 & 15.78377 & 62.78163 & 5.89 & 0.15 & 5.89 & 75.7 & 0.7 & -5.17 & 0.15 & 2.82 & 0.15\\
  28 & 15.78449 & 62.77241 & 0.96 & 0.23 & 0.93 & 54.8 & 6.8 & -0.32 & 0.23 & 0.91 & 0.23\\
  29 & 15.78481 & 62.79230 & 1.16 & 0.27 & 1.13 & 86.4 & 6.8 & -1.15 & 0.27 & 0.15 & 0.27\\
  30 & 15.78491 & 62.78920 & 3.26 & 0.2 & 3.25 & 71.8 & 1.7 & -2.62 & 0.2 & 1.93 & 0.2\\
  31 & 15.78585 & 62.79087 & 3.01 & 0.17 & 3.00 & 80.3 & 1.6 & -2.84 & 0.17 & 1.0 & 0.17\\
  32 & 15.78646 & 62.78017 & 4.84 & 0.19 & 4.84 & 73.5 & 1.1 & -4.06 & 0.19 & 2.64 & 0.18\\
  33 & 15.78806 & 62.79425 & 1.92 & 0.18 & 1.91 & 81.1 & 2.7 & -1.83 & 0.18 & 0.59 & 0.18\\
  34 & 15.79002 & 62.77905 & 2.7 & 0.22 & 2.69 & 73.9 & 2.4 & -2.28 & 0.22 & 1.44 & 0.22\\
  35 & 15.79058 & 62.78262 & 1.66 & 0.25 & 1.64 & 75.4 & 4.4 & -1.45 & 0.25 & 0.81 & 0.26\\
  36 & 15.79107 & 62.78132 & 2.18 & 0.23 & 2.17 & 77.7 & 3.0 & -1.98 & 0.23 & 0.9 & 0.23\\
  37 & 15.79161 & 62.79209 & 3.49 & 0.15 & 3.49 & 78.5 & 1.2 & -3.22 & 0.15 & 1.36 & 0.15\\
  38 & 15.79319 & 62.78464 & 3.06 & 0.16 & 3.06 & 84.7 & 1.5 & -3.0 & 0.16 & 0.56 & 0.16\\
  39 & 15.79691 & 62.79186 & 3.22 & 0.21 & 3.21 & 82.8 & 1.8 & -3.12 & 0.21 & 0.8 & 0.21\\
  40 & 15.79764 & 62.80114 & 2.56 & 0.22 & 2.55 & 91.2 & 2.5 & -2.56 & 0.22 & -0.11 & 0.22\\
  41 & 15.79963 & 62.78694 & 2.11 & 0.21 & 2.01 & 80.4 & 2.8 & -1.99 & 0.21 & 0.7 & 0.21\\
  42$^*$ & 15.80062 & 62.78256 & 0.42 & 0.29 & 0.30 & 106.1 & 19.5 & -0.36 & 0.29 & -0.22 & 0.29\\
  43 & 15.80225 & 62.80193 & 2.23 & 0.27 & 2.21 & 123.4 & 3.5 & -0.88 & 0.27 & -2.05 & 0.27\\
\enddata
\end{deluxetable*}
\subsection{Sky projection and distribution of polarization}\label{sec:3.1}
\begin{figure*}
	\plotone{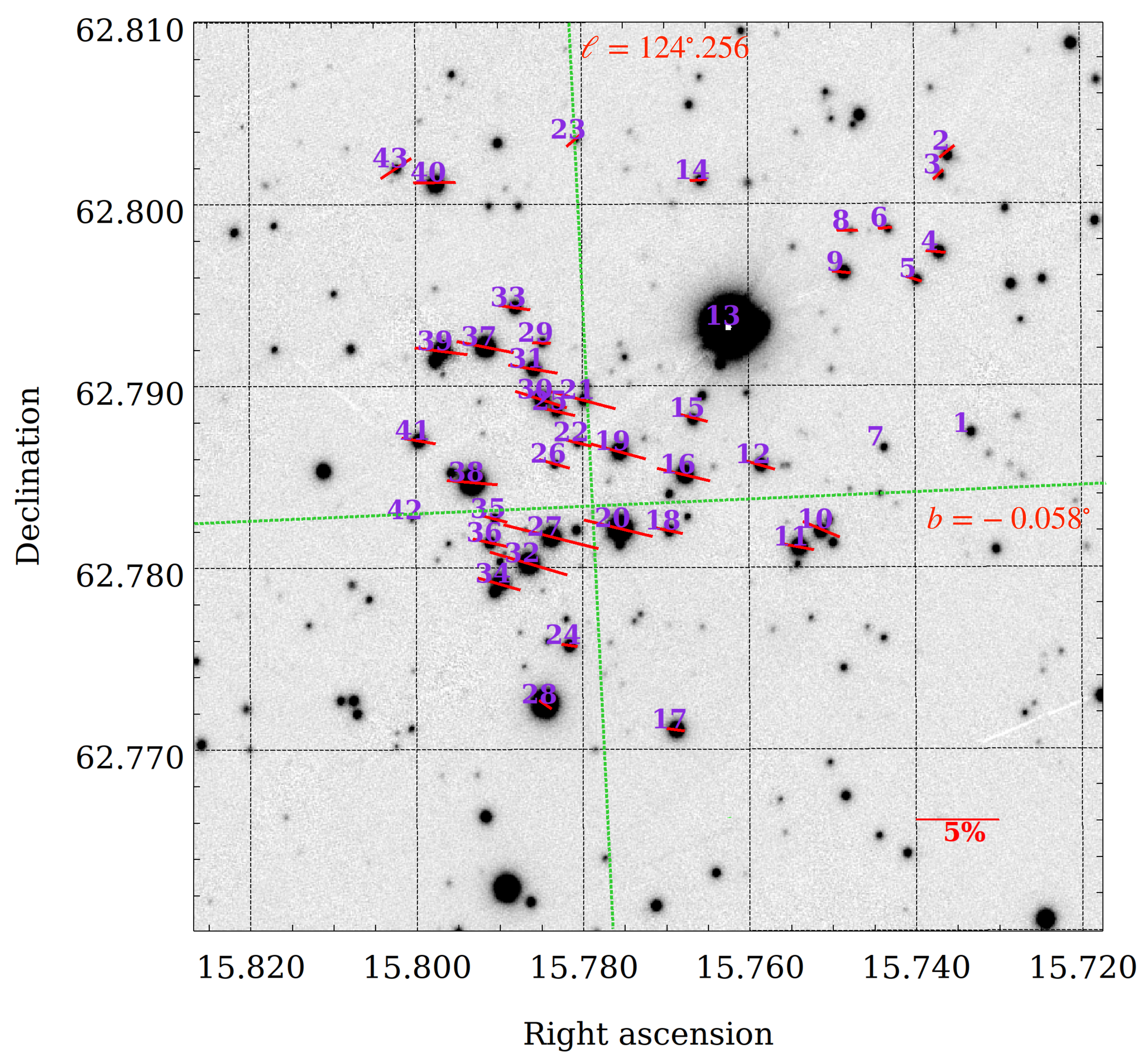}
    \caption{Polarization  measurement of 39 stars towards the Czernik 3 open cluster superimposed on $3^\prime \times 3^\prime$ \textit{g}-band image of PanSTARRS. Details are described in the text of Section \ref{txtlink}.
    }
    \label{fig:skyvectors}
\end{figure*}

 {The sky projection of the 39 stars is overlaid on the PanSTARRS g-band image of the cluster using the standard convention:  North at the top and East to the left in Figure \ref{fig:skyvectors}. The length of each red colored line segment is proportional to the degree of polarization and the orientation corresponds to the polarization angle measured from the North celestial pole increasing in direction of Right Ascension (RA). A reference line segment of 5\% polarization and $90^{\circ}$ polarization angle is drawn on the bottom-right side of the figure.
 The dotted grid lines in the figure correspond to  the equatorial coordinate system.  The orientation of the Galactic plane at the location of the cluster, $b=-0^\circ.058$, has a position angle of $\theta_{GP} \sim 91^\circ$ (angle between the constant latitude lines with north celestial pole increasing eastwards), and is shown by a green dashed line from left to right. The dashed green line from top to bottom of the figure corresponds to the galactic longitude $\ell=124^\circ.256$. These dashed lines intersect at the centre of the cluster ($\alpha_{J2000} = 01^h03^m06^s.9; \delta_{J2000}= 62^\circ47^\prime00^{\prime\prime}$) as determined by \cite{CZ3sharma2020}} \label{txtlink}. 

\begin{figure}[!h]
	\plotone{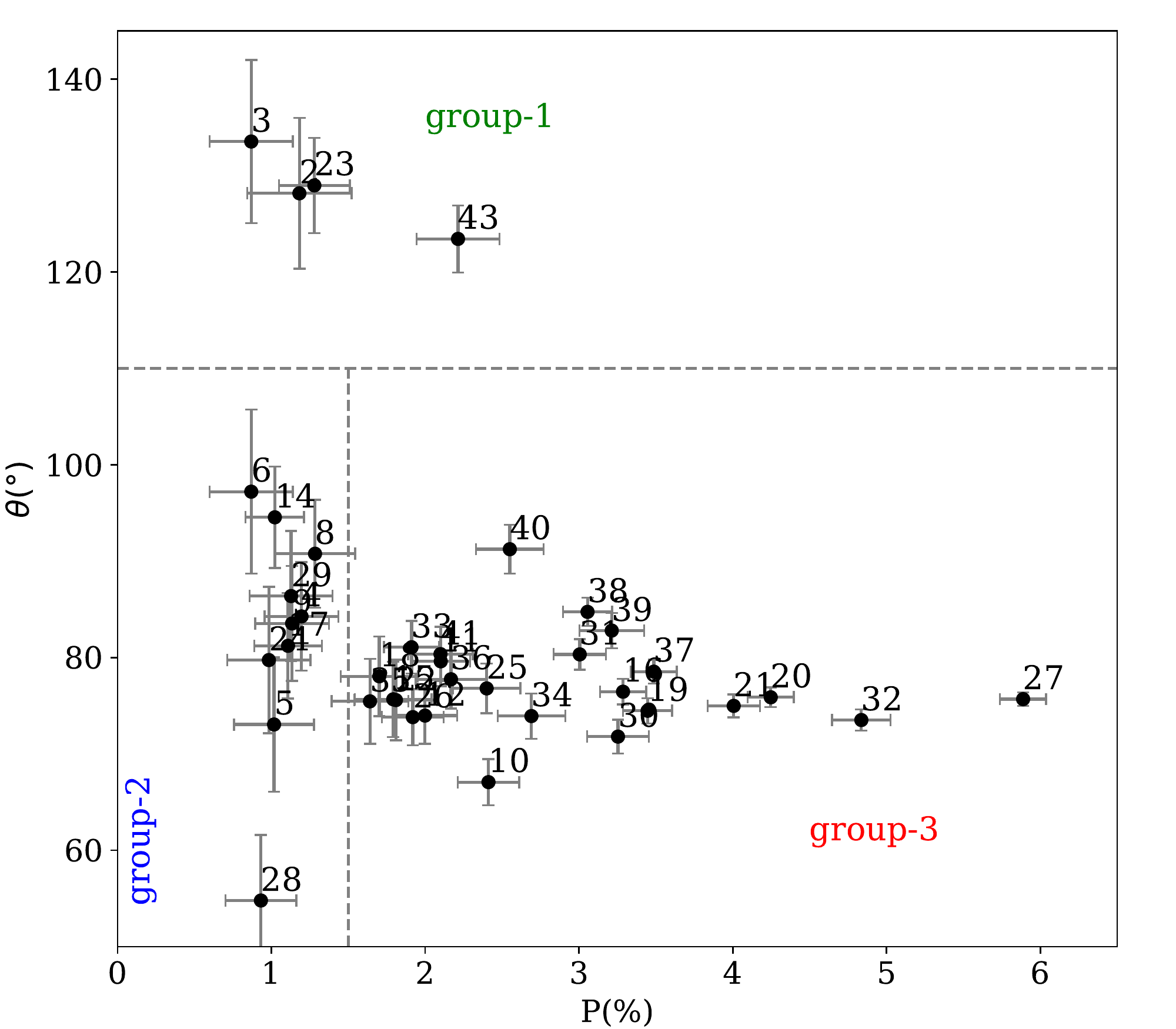}
    \caption{Polarization angle versus degree of polarization in \textit{Sloan i}-band of 39 stars towards Czernik 3 cluster.  The dashed lines are drawn to separate 3 groups in the observed sample - group-1 (green), group-2 (blue), and group-3(red).}
    \label{fig:pol_angle}
\end{figure}

Most of the stars in the field have polarization angle oriented at a small angle to the Galactic plane while there are a few stars showing a larger deviation.  The stars with larger deviation have slightly smaller degree of polarization (< 2.2 $\%$) and maybe foreground objects.  This is more clearly seen in the distribution of the polarization angle with the degree of polarization as shown in Figure \ref{fig:pol_angle}.
On the basis of visual inspection, we have divided the stars into three groups separated by dotted lines:
\begin{enumerate}
    \item The first group contains four stars \#2, \#3, \#23, and \#43 having degree of polarization ranging between $0.87\% < P < 2.21\%$ and polarization angle $123^\circ < \theta < 133^\circ$. The weighted mean polarization of this group is $1.39\%$ with dispersion of $0.50\%$. The average polarization angle is $127^\circ \pm 4^\circ$ showing a large deviation from the the galactic plane. This indicates that the stars in this group may be non-member/foreground stars.
    \item The second group consists of all the stars on the left side of the vertical dotted line marking $P < 1.5\%$ (Figure \ref{fig:pol_angle}). The stars present in this region have relatively smaller degree of polarization ($0.87\% < P < 1.28\%$) with weighted average value of $1.07\%$ and dispersion of 0.12$\%$ but large spread in polarization angle ($55^\circ < \theta < 97^\circ$). The weighted average orientation of this group ($83^\circ \pm 11^\circ$) is nearly parallel to the Galactic plane (having position angle of 91$^\circ$). These stars are distributed randomly in the outer regions of the core of the cluster (on the plane of the sky) and maybe foreground stars.
    \item The stars in the region with $P > 1.5\%$ (Figure \ref{fig:pol_angle}) forms the third group.  The degree of polarization and polarization angle of stars in this group are ranging from $1.64\% < P < 5.89\%$ and $67^\circ < \theta< 91^\circ$ respectively. The average polarization is $2.97\%$ with a dispersion of $1.04\%$.  A large range in degree of polarization has been observed in many other clusters like Trumpler 27 \citep{Trumpler27}, Hogg 22 and NGC 6204 \citep{Hogg22},  NGC 5749 \citep{NGC5749}, Berkeley 59 \citep{Bk59}. The large spread in the degree of polarization could be due to several reasons which are discussed in section \ref{4.2}. The weighted average polarization angle of the stars in this group  ($77^\circ \pm 5^\circ$) is deviated from the Galactic plane ($91^\circ$).
\end{enumerate}
 
\subsection{Distribution of polarization in the Stokes plane}
Polarization can be used as an efficient tool in determining the cluster membership even if the colors of the member stars and the field stars are the same. 
The light from an intrinsically unpolarized star becomes partially polarized after passing through the asymmetrical but aligned dust grains present in the dust clouds of the ISM. 
The degree of polarization depends on the column density of aligned dust grains along the line of sight. The stars foreground to the dust cloud will be less polarized as compared to the stars behind the dust layers. Also, the polarization angle of the two may differ from each other depending upon the direction of the local magnetic field.  It is expected that all the member stars are behind a common set of intervening clouds.  Therefore, their polarization properties are expected to be similar while the field stars may have slightly different properties.  The distribution of the stars in the Stokes $qu$-plane (with $q = P\cos2\theta$ and $u=P\sin2\theta$ ) is useful to distinguish the cluster members from the non-members. The members of the cluster are expected to group together in this plane while the field stars would show scattered distribution.  However, intrinsic polarization, rotation of polarization angle and patchy extinction can also give rise to scattered distribution in the $qu$-plane.
\begin{figure}[!h]
	\plotone{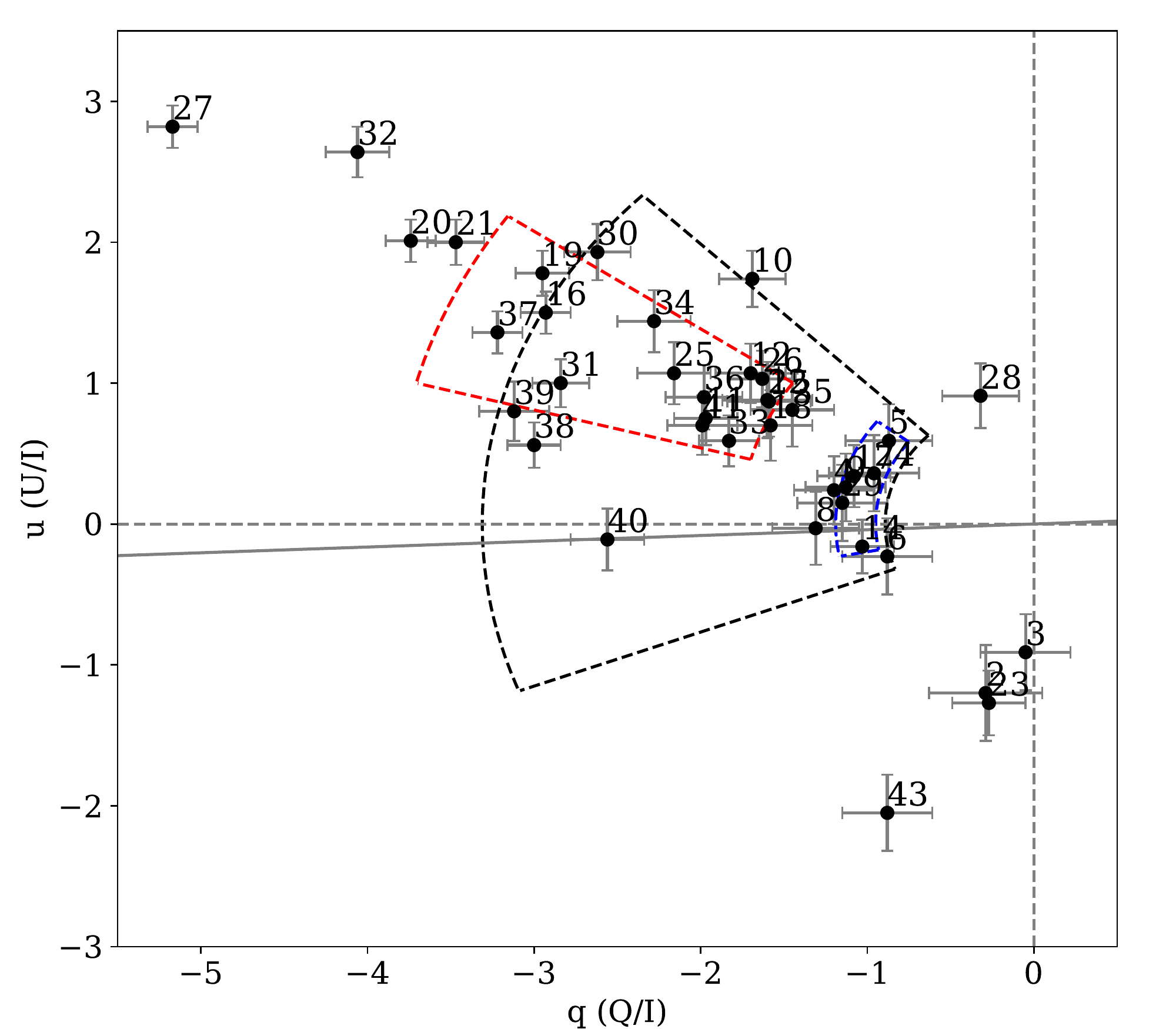}
    \caption{Distribution of the 39 observed stars in the $qu$ plane. Dotted line crossing at $q=u=0$ denotes the dustless solar neighbourhood and the gray solid line represents the Galactic parallel.
    Black dashed box is the 1 sigma boundary of all the stars observed  with mean polarization $2.22\% \pm 1.16\%$ and angle $84^\circ \pm 16^\circ$.
    1$\sigma$ box corresponding to mean degree of polarization ($0.93\% \pm 0.29\%$, $2.80\% \pm 1.04\%$) and polarization angles ($82^\circ \pm 11^\circ$, $77^\circ \pm 5^\circ$) of group-2 (region 2) and group-3 (region 3) of Figure \ref{fig:pol_angle} are shown as boxes of blue and red dotted lines respectively.}
    \label{fig:stokes}
\end{figure}

The normalized $q$ ($q = Q/I$) and $u$ ($u = U/I$) Stokes parameters are derived for all the observed stars and are plotted in Figure \ref{fig:stokes}. The cross-section of dashed straight lines at $q$ = $u$ = 0 represents the dustless solar neighbourhood and the solid gray line corresponds to the Galactic plane ($ \sim 91^\circ$). The stars, on average, show a scattered distribution in the $q-u$ plane, which is expected for distant clusters because of the contamination by large number of foreground and/or background field stars. 
This has also been seen in Stock 6,  NGC 1893 \citep{NGC1893}, and Berkeley 59 \citep{Bk59} clusters.  It is difficult to identify the grouping of member stars in such cases. We draw $1\sigma$ box (black dashed box in Figure \ref{fig:stokes}) with boundaries of mean $P\pm \sigma_P$ and $\theta \pm \sigma_{\theta}$ to elucidate the probable cluster members. All the stars within this box are considered as probable members and stars scattered away from the mean 1$\sigma$ box may have less membership probability. The $q$, $u$ for groups 2 and 3 (as described in previous section \ref{sec:3.1}) are ranging from $-1.31 \le q \le -0.32$; $-0.23 \le u \le 0.91$ and $-5.17 \le q \le -1.45$; $-0.11 \le u \le 2.82$ respectively. To check the distribution of stars in these regions, we plotted the mean $1\sigma$ box corresponding to these two groups in blue (region 2) and red dotted lines (region 3) respectively in Figure \ref{fig:stokes}. The blue dotted region is closer to the Sun and expected to consist of the foreground stars. While the member stars may be present in region 3. The stars outside the $1\sigma$ boxes are either field stars or having intrinsic component of polarization. 

The $q-u$ plot is also helpful to study the evolution of interstellar environment from the Sun to the cluster. The two regions are separated by a gap suggesting two dust layers present between the Sun and the cluster.  The first layer affects only the stars of region 2.  The stars of region 3 exhibit a cumulative effect of both dust layers in their polarization.  
 
 More details of the membership and dust distribution are discussed together with distance, extinction and astrometric information in section \ref{astrometry}.  
 
\section{Discussion}\label{sec:4}
In the first subsection (\ref{astrometry}) we redefine the cluster membership based on Gaia EDR3 astrometry.  Subsection \ref{4.2} 
discusses the polarization results of the 39 stars observed towards the core of Czernik 3 cluster in combination with different archival data.  Dust distribution towards the Czernik 3 cluster is described in subsection \ref{sec:4.3}.  The general trend of polarization and its implications regarding the dust distribution over a large spatial range is in subsection \ref{sec:4.4}.     
\subsection{Membership from Gaia EDR3 data}\label{astrometry}
The identification of member stars is important to determine the reliable cluster parameters like cluster radius, age and distance. In the past, photometry and polarization have been used to identify cluster members \citep[][etc.]{NGC1893, Bk59}. Detailed discussion of cluster membership for Czernik 3, on the basis of polarization and photometry, is described in Appendix \ref{ap:1}. However, with the availability of the Gaia data it is found that astrometry is the best method to determine cluster membership. We have calculated the success rate of membership determination from polarization as compared to the astrometric method and it comes out to be 58.3\% for Czernik 3 cluster.  Hence, we mainly focus on the cluster membership based on astrometric method using parallax and proper motion information.

\begin{figure}[!h]
	\plotone{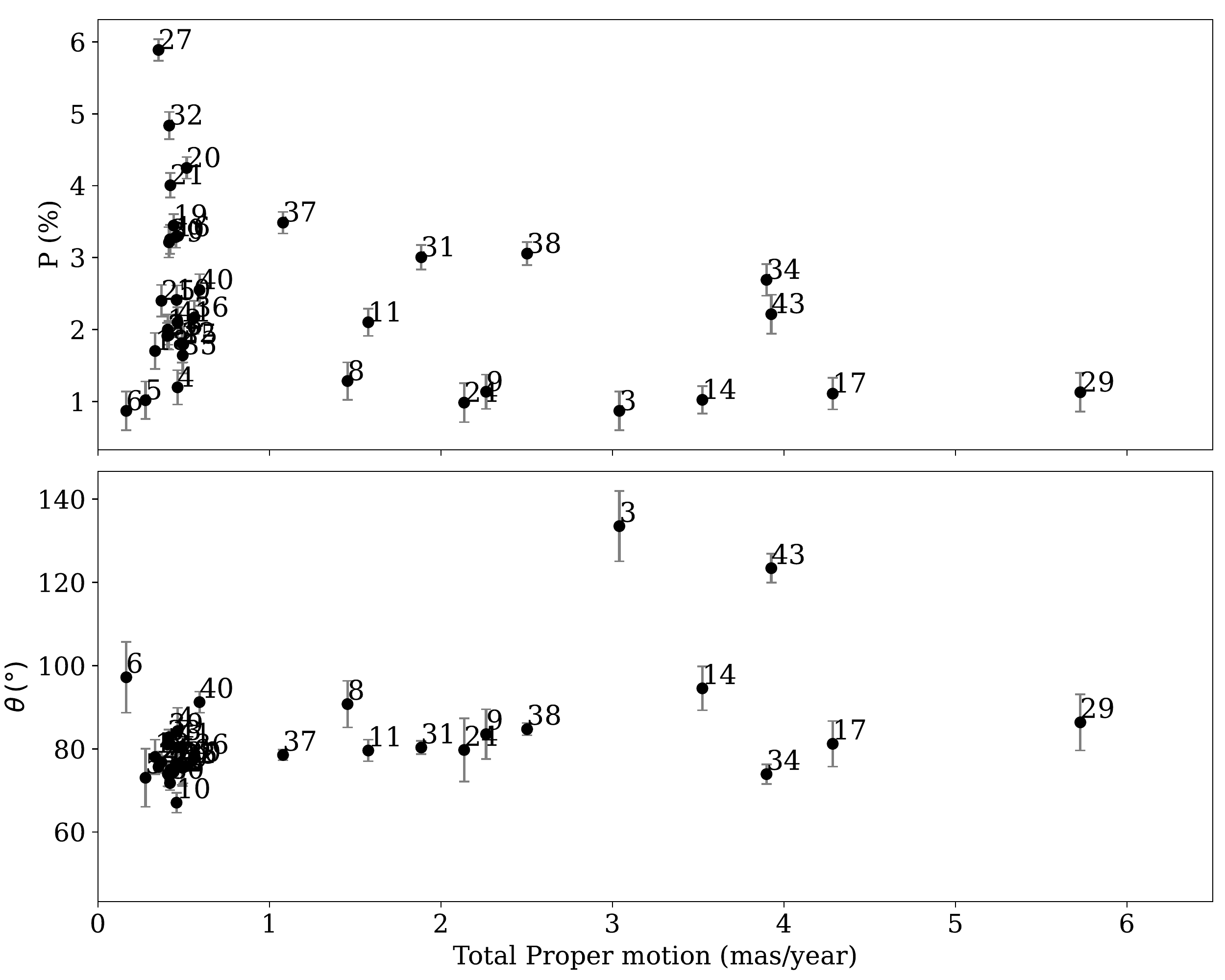}
    \caption{Observed polarization as a function of total proper motion of stars observed towards Czernik 3 cluster. Two stars (\#23 and \#28) are excluded from this plot  having PM $>$ 10 mas/yr.}
    \label{fig:propermotion}
\end{figure}

Figure \ref{fig:propermotion} shows the proper motion (PM) versus degree of polarization (upper panel) and polarization angle (lower panel) of 39 stars listed in Table \ref{tab:obs}. There is a prominent clump around proper motion of $\sim 0.5$ mas/year. Two stars \#23 and \#28 are not shown in the figure because they have proper motion $>$ 10 mas/year and are non-members from polarization and photometric methods also (see Appendix \ref{ap:1}). According to the proper motion distribution,  23 stars - \#4, \#5, \#6 \#10, \#12, \#15, \#16, \#18, \#19, \#20, \#21, \#22, \#25, \#26, \#27, \#30, \#32, \#33, \#35, \#36, \#39, \#40, and \#41 have higher membership probability than the other stars in the field showing scattered distribution in proper motion.\\
All the probable member stars (from Figure \ref{fig:propermotion})  except star \#6 and \#20 are also identified as members with probability $\ge 80\%$ (see column 4 of Table \ref{tab:mem}) using proper motion and parallax information of Gaia DR2 in UPMASK membership assignment code by \citet{gaiaDR2OC}. On the other hand \citet{CZ3sharma2020} have selected the stars with high probability ($\ge 90\%$) of membership based on frequency distribution of member stars and field stars using Gaia DR2 proper motion information and have only 17 stars common with probable members from Figure \ref{fig:propermotion} (see column 5 of Table \ref{tab:mem}).  This has resulted in discrepancies of membership for individual stars between the two studies, as can be seen in the Table \ref{tab:mem}. 

\begin{figure}[!h]
	\plotone{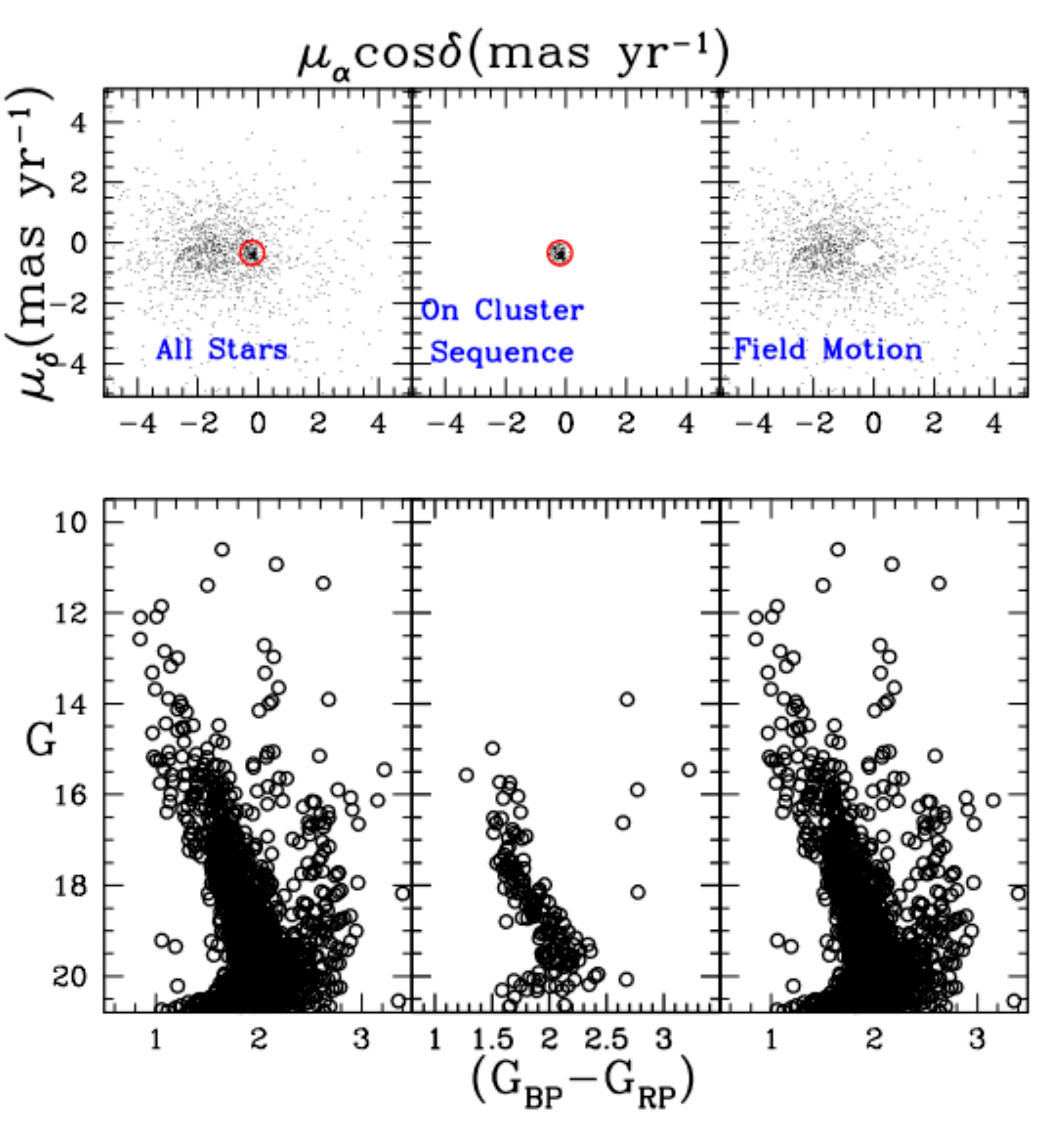}
    \caption{(Top row) vector point diagrams for cluster Czernik 3. (Bottom row) $G$ versus $(G_{BP}-G_{RP})$ color-magnitude diagrams. (Bottom left panel), The entire sample. (Bottom  middle panel) Stars within the circle of $0.5~ mas~ yr^{-1}$ radius. (Bottom right panel) Probable background/foreground field stars in the direction of the cluster Czernik 3.}
    \label{fig:vpd}
\end{figure}

We have used the latest Gaia proper motion (PM) data (from Gaia EDR3) to redefine the membership probability. PMs, $\mu_{\alpha} \cos{\delta}$ and $\mu_{\delta}$ are plotted as vector point diagrams (VPDs) in the top row of Figure \ref{fig:vpd} to see the distribution of cluster members and field  stars towards the region of Czernik 3. The bottom row of this figure show the corresponding $G$ versus $(G_{BP}-G_{RP})$ color magnitude diagrams (CMDs). The left panel in the CMD shows all the stars present within five arcmin radius around Czernik 3, while the middle and right panels show the possible cluster members and field stars, respectively. By visual inspection, we define the cluster center and radius in the VPD iteratively so as to reduce the contamination by the field stars while keeping a possible number of faint stars in the cluster sequence (middle panel of the CMD). 
A circle of 0.5 mas yr$^{-1}$  around the center of the member stars distribution in the VPDs indicates our membership zone. We can see in the middle-bottom panel that the cluster's main sequence is separated. 

To identify the cluster members quantitatively, we have estimated the membership probabilities of the stars following the method given by \citet{BN98}.
Many authors have previously used this method \citep{bellini2009,CZ3sharma2020,DB21,DPS21}. Two distribution functions ($\phi_{c}^{\nu}$) and ($\phi_f^{\nu}$) for cluster and field stars are constructed for a particular i$^{th}$ star, which are given as follows:
\begin{eqnarray}
    \phi_c^{\nu} &=&\frac{exp\left\{ -\frac{1}{2}\left[\frac{(\mu_{xi} - \mu_{xc})^2}{\sigma_c^2 + \epsilon_{xi}^2 } + \frac{(\mu_{yi} - \mu_{yc})^2}{\sigma_c^2 + \epsilon_{yi}^2}\right] \right\}}{2\pi\sqrt{{(\sigma_c^2 + \epsilon_{xi}^2 )} {(\sigma_c^2 + \epsilon_{yi}^2 )}}} \\[5pt]
\phi_f^{\nu}&=&\frac{1}{2\pi\sqrt{(1-\gamma^2)}\sqrt{{ (\sigma_{xf}^2 + \epsilon_{xi}^2 )}{(\sigma_{yf}^2 + \epsilon_{yi}^2 )}}} \times \nonumber\\& &  exp\Bigg[ -\frac{1}{2(1-\gamma^2)} \Big( \frac{(\mu_{xi} - \mu_{xf})^2}{\sigma_{xf}^2 + \epsilon_{xi}^2}- \nonumber\\& &\frac{2\gamma(\mu_{xi} - \mu_{xf})(\mu_{yi} - \mu_{yf})} {\sqrt{(\sigma_{xf}^2 + \epsilon_{xi}^2 ) (\sigma_{yf}^2 + \epsilon_{yi}^2 )}}+ \frac{(\mu_{yi} - \mu_{yf})^2}{\sigma_{yf}^2 + \epsilon_{yi}^2}\Big)\Bigg]
\end{eqnarray}
where, ($\mu_{xi}$, $\mu_{yi}$) are the PMs of $i^{th}$ star. The PM errors are represented by ($\epsilon_{xi}$, $\epsilon_{yi}$). The cluster's PM center, given by ($\mu_{xc}$, $\mu_{yc}$) and ($\mu_{xf}$, $\mu_{yf}$) represents the center of field PM values. The intrinsic PM dispersion for the cluster stars is denoted by $\sigma_c$, whereas $\sigma_{xf}$ and $\sigma_{yf}$ provide the intrinsic PM dispersion's for the field populations. The correlation coefficient $\gamma$ is calculated as:
\begin{equation}
\gamma = \frac{(\mu_{xi} - \mu_{xf})(\mu_{yi} - \mu_{yf})}{\sigma_{xf}\sigma_{yf}}
\end{equation}

In order to estimate probability of Czernik 3 star, we used only stars having PM errors better than $\sim$1 mas~yr$^{-1}$.  We found a clear bunch of stars at $\mu_{xc}$ = $-$0.22 mas~yr$^{-1}$, $\mu_{yc}$ = $-$0.32 mas~yr$^{-1}$ and in the circular region with a radius of 0.5 mas~yr$^{-1}$ (Figure \ref{fig:vpd}).  We estimated dispersion ($\sigma_c$) in PMs as 0.06 mas~yr$^{-1}$ by using cluster distance of 3.5 kpc \citep{CZ3sharma2020} and the radial velocity dispersion of 1 km $s^{-1}$ for open star clusters \citep{Girard89}. For field region stars, we have estimated ($\mu_{xf}$, $\mu_{yf}$) = ($-$0.95, $-$0.55) mas yr$^{-1}$ and ($\sigma_{xf}$, $\sigma_{yf}$) = (6.5, 4.4) mas yr$^{-1}$.

In consideration of the normalized numbers of cluster and field stars as $n_{c}$ and $n_{f}$ respectively (i.e., $n_c + n_f = 1$), the total distribution function can be estimated as 
\begin{equation}\label{9}
\phi = (n_{c}~\times~\phi_c^{\nu}) + (n_f~\times~\phi_f^{\nu}), 
\end{equation}
Finally, the membership probability for the $i^{th}$ star is given by
\begin{equation}\label{10}
P_{\mu}(i) = \frac{\phi_{c}(i)}{\phi(i)}. \\
\end{equation}

We obtain 72 cluster members in Czernik 3 cluster with membership probability higher than $90\%$ and $G\le18.5$ mag. 

The stellar density of the cluster region is affected by the presence of field region stars. So, we have calculated the effectiveness of membership determination for Czernik 3 using the formula given in eq. (\ref{11}) \citep{shao96}:
\begin{equation}\label{11}
    E=1-\frac{N\times\Sigma[P_{i}(1-P_{i})]}{\Sigma P_{i}\Sigma(1-P_{i})}
\end{equation}

where, $N$ is the total number of cluster members and $P_{i}, $ indicates the probability of $i^{th}$ star of the cluster. The effectiveness $(E)$ value is obtained as $\sim$ 0.70 for Czernik 3. \citet{shao96} show that the effectiveness of membership determination of 43 open clusters ranges from 0.20 to 0.90 with the peak value of 0.55. Our estimated value is on the higher side but lies within the above limits.
The mean distance of the Czernik 3 cluster is calculated using the distance information \citep{bailer2021} of the member stars present in the core \citep[$0.5^\prime$,][]{CZ3sharma2020} of the cluster. The average distance of the core members comes out to be $3.6 \pm 0.8$ kpc, which matches well with the distance determined by \citet{CZ3sharma2020} using Gaia DR2 data. The membership probability of the observed stars is given in column 6 of Table \ref{tab:mem}. 

A similar procedure is adopted to redefine the membership of the clusters IC 1805, NGC 654, Berkeley 59, NGC 457, and Alessi 1 using Gaia EDR3 data. From the revised cluster membership, the mean distance of each cluster is constrained to be  2.55 $\pm$ 0.89, 2.89 $\pm$ 0.72, 1.21 $\pm$ 0.73, 2.87 $\pm$ 0.72, and 0.72 $\pm$ 0.17 kpc respectively. These  values closely match with the average distance of the respective clusters in \citet{gaiaDR2OC} (see Table \ref{tab:clusters}).

\begin{deluxetable*}{llcccc}[!htb]
\tablecaption{Czernik 3 cluster members with membership probabilities from literature (column 4 and 5) and our study (column 6) with IDs (column 1) and polarization (column 2 and 3). Asterisk (*) and double asterisk (**) IDs represents the member YSO candidates selected from \citet{YSOs} and Q-parameter method discussed in the section \ref{sec:4.2.3a}\label{tab:mem}}

\tablewidth{0pt}
\tablehead{
\colhead{ID} & \colhead{$P \pm \epsilon_{P} $} & \colhead{$\theta \pm \epsilon_{\theta}$}  & \colhead{Membership probability ($\%$)}  & \colhead{Membership probability ($\%$)}  & \colhead{Membership probability ($\%$) }\\
 &   \colhead{($\%$)} & \colhead{($^\circ$)} & \colhead{\citep{gaiaDR2OC}} & \colhead{\citep{CZ3sharma2020}} & \colhead{(our study)}\
}
\startdata
4 & $1.20 \pm 0.24$ & $84\pm6$ &   100 & - & 99.61  \\
$5^{**}$ & $1.02 \pm 0.26$ & $73\pm7$ &   90 & - & 99.93 \\
6 & $0.87 \pm 0.27$ & $97\pm8$ &   - & 90 & 99.72 \\
$10^*$ & $2.41\pm 0.20$ & $67\pm2$ & 100 & 100 & 99.9 \\
$12^{**}$ & $2.00 \pm 0.21$ & $74\pm3$ & 80 & 99 & 99.43 \\
15 & $1.79 \pm 0.25$ & $75\pm4$ &  100 & 99 & 99.93 \\
16 & $3.29 \pm 0.15$ & $76\pm1$ &  100 & 100 & 99.97 \\
18 & $1.70 \pm 0.25$ & $78\pm4$ &  100 & 100 & 99.94 \\
$19^*$ & $3.45 \pm 0.16$ & $74\pm1$ & 100 & 100 & 99.98 \\
20 & $4.25 \pm 0.15$ & $76\pm1$ & - & 100 & 99.66 \\
21 & $4.01 \pm 0.17$ & $75\pm1$ & 80 & - & 99.76 \\
22 & $1.81 \pm 0.27$ & $76\pm4$ &   100 & 100 & 99.88 \\
$25^{**}$ & $2.40 \pm 0.22$ & $77\pm3$ &  100 & 100 & 99.95 \\
26 & $1.92 \pm 0.20$ & $74\pm3$ &   80 & - & 99.96 \\
27 & $5.89 \pm 0.15 $ & $76\pm1$ & 100 & 99 & 99.83 \\ 
$30^{**}$ & $3.25 \pm 0.20$ & $72\pm2$ & 100 & 100 & 99.97 \\
32 & $4.84 \pm 0.19$ & $73\pm1$ & 100 & 100 & 99.97 \\
33 & $1.91\pm 0.18$ & $81\pm3$ & 100 & - & 99.64 \\
35 & $1.64 \pm 0.25$ & $75\pm4$ &  90 & 99 & 99.89 \\
36 & $2.17 \pm 0.23$ & $78\pm3$ &  100 & 100 & 99.80 \\
39 & $3.21 \pm 0.21$ & $83\pm2$ &   100 & 100 & 99.98 \\
$40^*$ & $2.55 \pm 0.22$ & $91\pm2$ &  100 & - & 98.52\\
41 & $2.01 \pm 0.21$ & $80 \pm 3$&  100 & 99 & 99.96 \\
\enddata
\end{deluxetable*}
\subsection{On the variation of polarization over the Czernik 3 cluster}\label{4.2}
In this section we discuss the possible reasons for the large range in polarization observed over the stars of Czernik 3 cluster.
\subsubsection{Patchy dust distribution from WISE W4 image}\label{4.2.1a}
\begin{figure*}[!h]
	\plotone{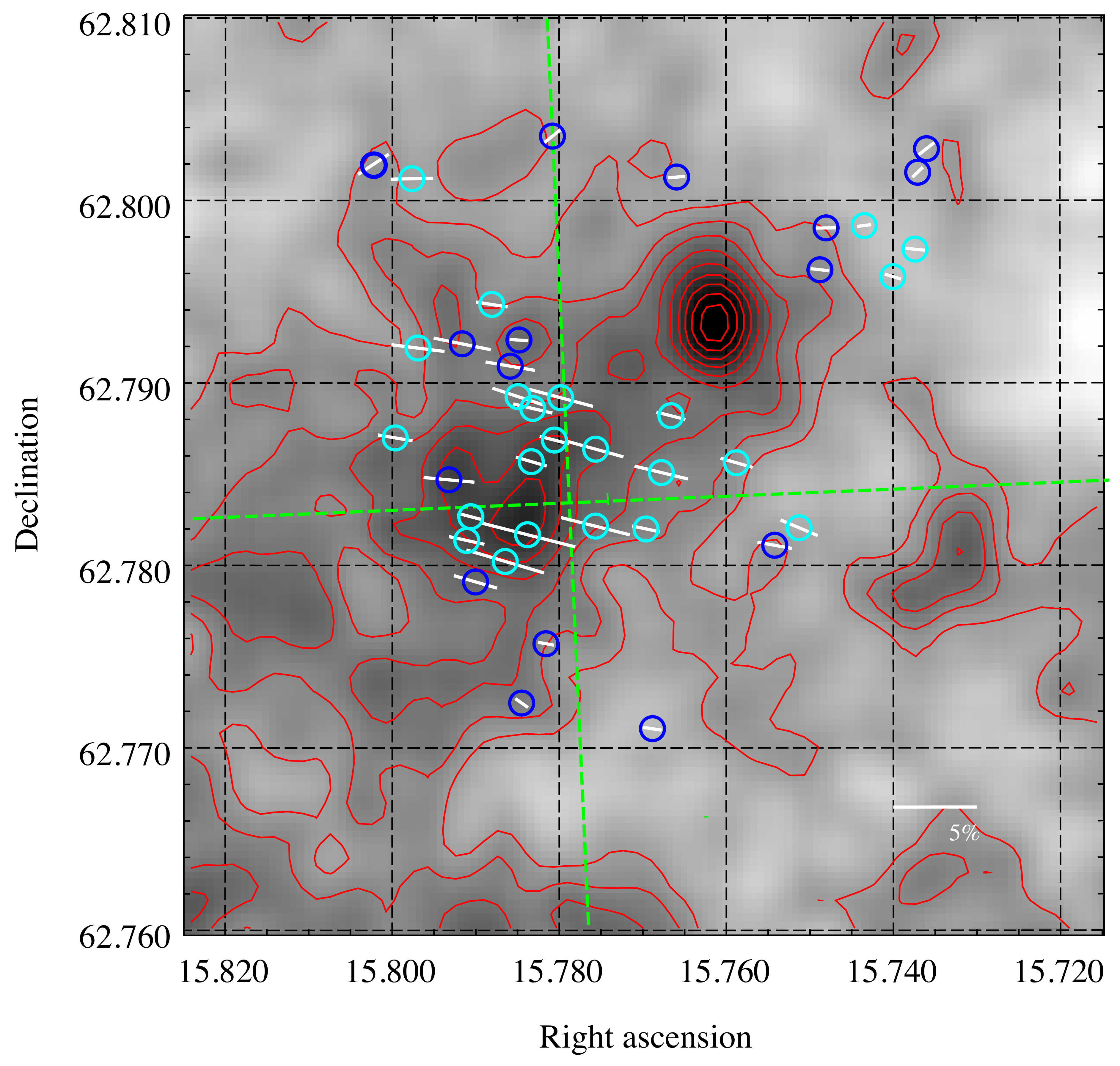}
    \caption{WISE W4 integrated intensity map of $3^{\prime}$ region around Czernik 3 with 10 contour levels between 113.02-113.508 DN at the interval of 0.054 DN. Polarization measurements are denoted by white colored lines. The cyan and blue circles centered at the star location represent the members and non-member stars respectively. The green dashed lines correspond to the Galactic coordinate grid passing through the center of the cluster.}
\label{fig:WISE}
\end{figure*}

Large dispersion in polarization values (Figure \ref{fig:pol_angle}) 
could be due to the patchy distribution of dust along the line of sight towards  Czernik 3 cluster, or due to the presence of dust within the cluster. This is further examined by analysing  $22\mu m$, WISE W4-band image\footnote{https://irsa.ipac.caltech.edu/applications/wise/}. Figure \ref{fig:WISE} shows the  spatial distribution of dust (integrated emission along the line of sight) traced by WISE W4 band in $3^\prime$~field around Czernik 3. 
The red color contours represent the equal flux density points. Patchiness is well observed from the contour levels varying from 113.02 DN to 113.508 DN with interval of 0.054 DN.  The uncertainty levels are $\sim 0.04$ DN. This implies that the dust is not distributed uniformly. It may be constant in small spatial areas but is highly variable across the face of the cluster. For example the central region of the cluster has higher flux density. Higher polarization in the same region is also observed, as seen by the length of the white colored polarization measurements centered at the star location.  This is further discussed in section \ref{sec:4.2.3}. Cyan and blue colored open circles in the figure denote the member and non-member stars. The patchy distribution has also been observed earlier in NGC 6823 open cluster embedded in the highly extincted region towards the inner Galaxy \citep{NGC6823}. 

Stars \#4, \#5, and \#6 have low polarization $1.20\% \pm 0.24\%$, $1.02\% \pm 0.26\%$, and $0.87\% \pm 0.27\%$ with large uncertainty in the polarization angle ($84^\circ.3 \pm 5^\circ.6$, $73^\circ.0 \pm 7^\circ.0$, and $97^\circ \pm 8^\circ$). Even though these stars are present in $1\sigma$ box of group-2 (Figure \ref{fig:stokes}), their proper motion data categorizes them as member stars with high membership probability ($> 99.5\%$). The low value of the polarization can be explained by the lesser dust column density traversed by the light coming from these stars, as seen by the contour level being $\le 113.02$ DN.

On the contrary, the core region stars \#21, \#27 and \#32 show large degree of polarization ($4.01 \pm 0.17\%$, $5.89 \pm 0.15$ and $4.84 \pm 0.19 \%$) because of higher dust density at central region as compared to peripheral region. The proper motion and polarization angle ($75^\circ.0 \pm 1^\circ.2$, $75.7^\circ \pm 0^\circ.7$ and $73.5 \pm 1^\circ.1$) of these stars are similar to the member stars but lying outside region 3 in $qu$-plot (Figure \ref{fig:stokes}), which may be due to non-uniform dust distribution.

\subsubsection{Polarization efficiency}
\begin{figure}[!h]
	\plotone{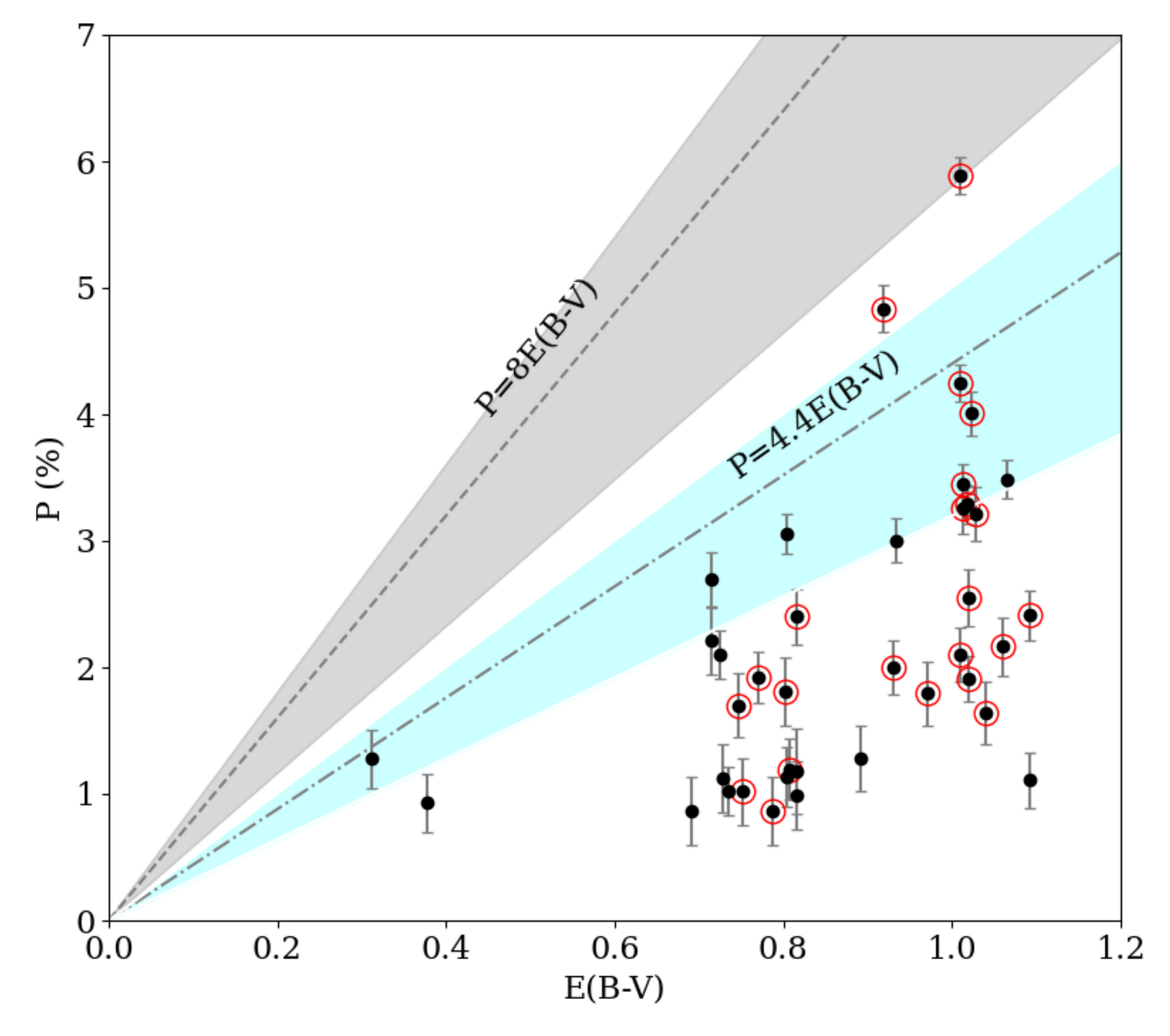}
    \caption{Polarization efficiency towards Czernik 3 cluster. Two gray lines: dashed and dashed-dotted represent the empirical upper limit and the average value for the diffuse ISM respectively in \textit{Sloan} \textit{i}-band using $\lambda_{max}$ = 0.55 $\micron$ and K = 1.15.See the text for further details of the shaded regions.}
    \label{fig:efficiency}
\end{figure}

The degree of polarization produced for a given amount of extinction is referred to as polarization efficiency of the intervening dust grains. It is known that the polarization efficiency of the ISM is non-uniform.  This can be seen from large scatter in Figure \ref{fig:efficiency}, which shows the relation between color excess and i-band polarization for the stars observed by us along the line of sight of Czernik 3 cluster. The efficiency depends on the degree of alignment of the dust grains with the magnetic field and also on the amount of depolarization due to radiation traversing more than one cloud with different magnetic field directions. The empirical upper limit \citep{Hiltner1956} for maximum polarization efficiency of diffuse ISM at visual wavelengths assuming $R_V = 3.1$ \citep{Seaton} is given by
\begin{equation}\label{12}
    \frac{P_V}{E_{B-V}} < 9.0\:\%\, mag^{-1}
\end{equation}

The polarization efficiency in \textit{Sloan i}-band ($\lambda_{eff}= 0.767\mu m$) is determined using Serkowski's law for interstellar polarization by assuming maximum polarization at average wavelength $\lambda_{max} \sim 0.55\mu m$ with K= 1.15 \citep{Whittet2003} and is given by  eq. (\ref{13}) (dashed line in Figure \ref{fig:efficiency}).
\begin{equation}\label{13}
    \frac{P_{i}}{E(B-V)} < 8.0\% mag^{-1}
\end{equation}
However, the $\lambda_{max}$ varies from star to star and has a typical range of $0.3 \mu m - 0.8\mu m$ \citep{Whittet2003}. Many studies \citep[e.g,][]{Wilking1980, Wilking1982, Clayton1995,martin} have shown that the $\lambda_{max}$ and K are linearly correlated by the Wilking law:
\begin{equation} \label{eq:wikling}
    K = C_1~\lambda_{max} + C_2
\end{equation}
Where, the constants $C_1$ and $C_2$ are given by $1.66 \pm 0.01$ $\mu$m$^{-1}$ and $0.01 \pm 0.05$ in visible to near infrared (VIR) wavelength regime ($0.35 \mu m \le \lambda \le 2.2 \mu m$) \citep{whittet1992dust}. In order to explore the effect of polarization efficiency in the \textit{Sloan i}-band, we have considered  $0.3 \mu m \le \lambda_{max} \le 0.8 \mu m$ and eq. (\ref{eq:wikling}) for the K values, in the Serkowski law of interstellar polarization. The resulting broad range of empirical relations cover an area demarcated by light grey colored region in the Figure \ref{fig:efficiency}.
The polarization efficiency of the diffuse ISM, in general, follows the mean relation $P_{max} = 5~E(B-V)$ \citep{Serkowski1975}. The corresponding \textit{Sloan} \textit{i}-filter polarization as a function of E(B-V) is calculated to be $P_{i}=4.4~E(B-V)$ and it is shown by dash-dotted line (with $\lambda_{max}$ = 0.55 $\mu$m, K=1.15) and cyan shaded region (corresponding to the range in $\lambda_{max}$) in the same figure. The observed polarization efficiency towards the cluster is less than the empirical upper limit defined by the light grey shaded region described earlier. A few stars (member as well as non-member) are showing efficiency more than the general diffuse ISM considering varying $\lambda_{max}$ as seen from their appearance above the lower limit of the cyan shaded region. Out of these,  two stars (\#27 and \#32) are lying above the maximum of the cyan shaded region, indicating different polarization behaviour as compared to the diffuse ISM. These two stars are near the center of the cluster, where there is an excess of dust (see the discussion in section \ref{4.2.1a}).  
It is also possible that they are both intrinsically polarized (see the discussion in Section \ref{sec:4.2.3a}) for example, \#27 is an evolved (probable AGB) star where one may expect intrinsic polarization due to an asymmetric dust shell.

\subsubsection{Intrinsic polarization}\label{sec:4.2.3a}
Stars having circumstellar dust, can show intrinsic polarization which may be different from the interstellar polarization. Young stellar objects and evolved stars are the possible candidates to show intrinsic polarization. Recently, \citet{YSOs} classified stars into four categories, i.e., extincted main sequence stars, evolved stars, extra-galactic sources and, young stellar objects using machine learning techniques by considering Gaia DR2 and ALLWISE data. We have found entries for 10 of our observed stars in this catalog\footnote{https://vizier.u-strasbg.fr/viz-bin/VizieR?-source=II/360}.  The probabilities of each category  with their errors are listed in Table \ref{tab: 3}.
We have opted to use their probability columns which do not consider the W3 and W4 band fluxes since the `R' parameter (first column in the Table \ref{tab: 3}), which represents the probability for W3 and W4 detections to be real, is less than 75\% for all 10 stars.
From the Table \ref{tab: 3}, 7 stars (\#3, \#10, \#17, \#19, \#23, \#28 and \#40) are highly probable YSO candidates, out of which \#10 \#19 and \#40 (with underlined ID) are cluster members and stars \#27 and \#38 have high probability of being evolved stars.

\begin{deluxetable*}{cccccccccc}
\tablecaption{Probability of source being Evolved (SE), Extra-galactic (SEG), Main sequence S(MS), or YSOs (SY) with their errors from the catalog by \citet{YSOs}.
R - probability that the W3 and W4 detections are real. The probabilities are written in bold for most probable category and the IDs of the member stars are underlined.\label{tab: 3} }
    
\tablehead{
\colhead{ID} & \colhead{R} & \colhead{SE}  & \colhead{$\epsilon_{SE}$}  & \colhead{SEG}  & \colhead{$\epsilon_{SEG}$ } & \colhead{SMS} & \colhead{$\epsilon_{SMS}$} & \colhead{SY} & \colhead{$\epsilon_{SY}$}
}
\startdata
3 & 0.378 & 0.0784 & 0.04382 & 0.1492 & 0.06497 & 2.0E-4 & 6.3E-4 & \textbf{0.7722} & 0.09265\\
  \underline{10} & 0.448 & 0.0768 & 0.03213 & 0.2426 & 0.11013 & 0.0236 & 0.06561 & \textbf{0.657} & 0.10294\\
  17 & 0.438 & 0.0572 & 0.0351 & 0.0664 & 0.04666 & 0.0094 & 0.01276 & \textbf{0.867} & 0.07939\\
  \underline{19} & 0.278 & 0.0704 & 0.0182 & 0.0848 & 0.02443 & 0.0048 & 0.00492 &\textbf{ 0.84} & 0.03307\\
  23 & 0.43 & 0.1234 & 0.06507 & 0.0482 & 0.01853 & 0.0012 & 0.0038 & \textbf{0.8272} & 0.07272\\
  \underline{27} & 0.646 & \textbf{0.6674} & 0.08052 & 0.0076 & 0.0044 & 0.0112 & 0.01237 & 0.3138 & 0.07726\\
  28 & 0.348 & 0.2232 & 0.03774 & 0.0172 & 0.01084 & 0.0336 & 0.01622 & \textbf{0.726} & 0.04233\\
  38 & 0.65 & \textbf{0.9118} & 0.04445 & 0.0224 & 0.02948 & 0.0084 & 0.00858 & 0.0574 & 0.01718\\
  \underline{40} & 0.45 & 0.101 & 0.03923 & 0.0576 & 0.02965 & 0.0058 & 0.00577 & \textbf{0.8356} & 0.04639\\
\enddata
\end{deluxetable*}
YSO candidates can also be identified using reddening parameter Q. The stars are considered as a candidate young stellar object if the reddening-free  parameter Q becomes less than $-0.05$ \citep{buckner2013properties}. The Q value can be estimated for a star using the VVV photometric magnitude relationship given by 
\begin{equation}
    Q = (J-H)-1.55\times(H-K)
\end{equation}
We used the 2MASS photometric values for our stars transformed to VVV using transformation equations of 1.5 version of CASU photometry \citep{2MASSVVV} to obtain the Q parameter.
We find 5 candidate YSOs (\#5, \#12, \#24, \#25, \#30) using the above relation of which \#5, \#12, \#25, and \#30 are cluster members.  

Thus, in total 7 member YSOs candidates have been detected in Czernik 3. The intrinsic polarization due to the circumstellar disk can be a cause of scattered position in the $qu$-plane for some of these stars.  These stars are marked with asterisk and double asterisk in the ID column of the Table \ref{tab:mem}.
\subsubsection{Spatial variation of polarization}\label{sec:4.2.3}
\begin{figure}[!htb]
	\plotone{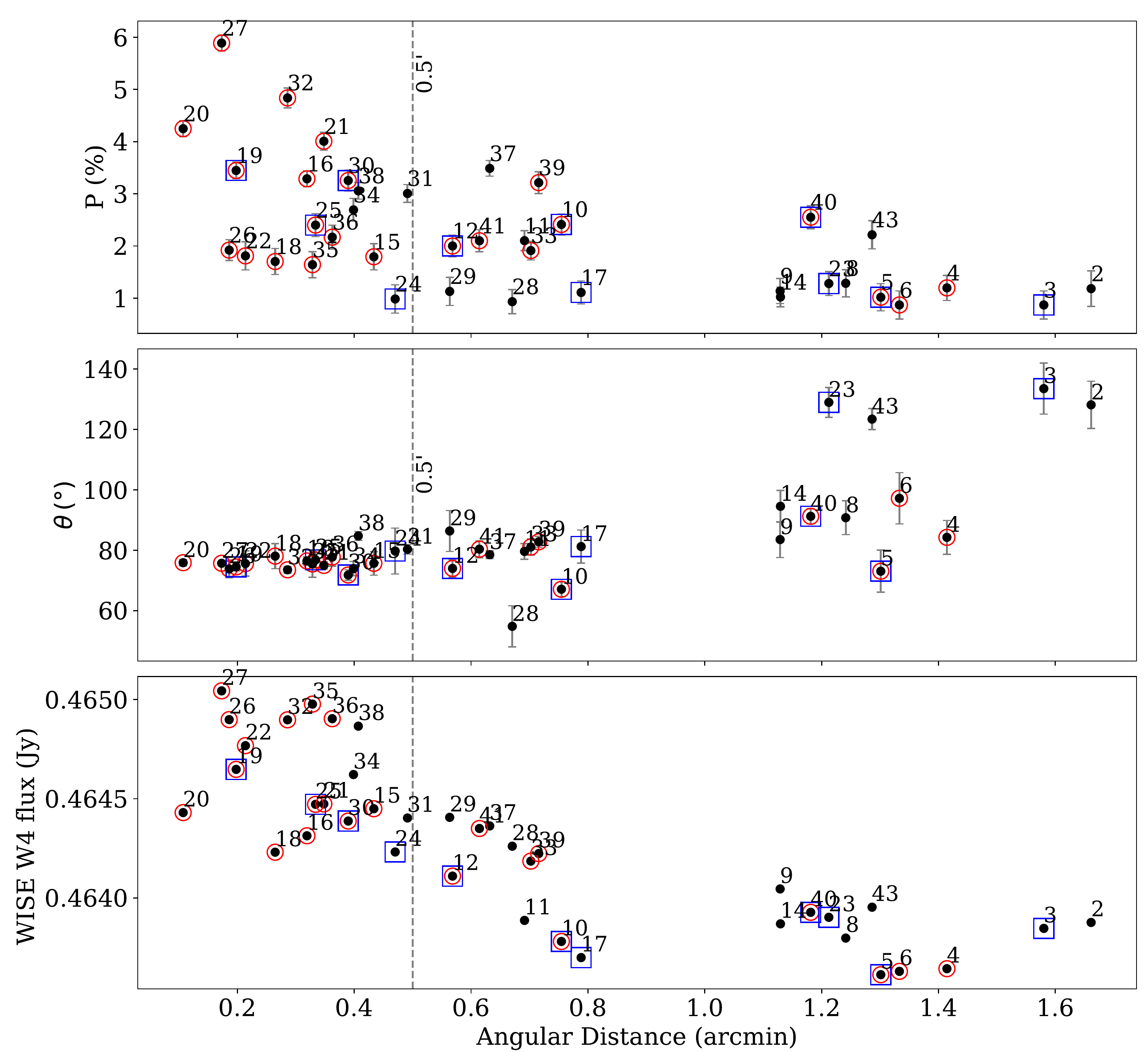}
    \caption{Variation of degree of polarization, polarization angle and WISE flux counts of stars towards Czernik 3, with angular distance from center of cluster. Member stars and YSO candidates are denoted in red circle and blue squares respectively.}
    \label{fig:angDistSh}
\end{figure}

First two panels of Figure \ref{fig:angDistSh}  show the variation of degree of polarization and polarization angle with angular distance from the center of the cluster. 
The angular distance of a star having equatorial co-ordinates ($\alpha, \delta$) from the center of the cluster ($\alpha_{c}, \delta_{c}$) is given by the eq. (\ref{15}).
\begin{equation}\label{15}
     d = 2 sin^{-1}\left\{\sqrt{sin^2\frac{|\delta-\delta_{c}|}{2} + cos\delta\; cos\delta_{c}\; sin^2 \frac{|\alpha-\alpha_{c}|}{2}}\right\}
\end{equation}
The third panel represents the WISE W4 flux (in Jy) as a function of angular distance. This flux is obtained by using aperture photometry on the position of the stars using a fixed aperture. A  decrease is seen in the  WISE flux as well as degree of polarization radially outwards. The polarization angle exhibits a very small dispersion in the core region (radius of $0.5^{\prime}$ \citep{CZ3sharma2020} - marked by dashed line) while the dispersion increases in the outer directions. This suggests that the dust is more concentrated towards the center and the density decreases with the radial distance from the cluster center.  The small dispersion in polarization angle with increased degree of polarization also indicates that the dust in the cluster is oriented in the same direction as the dust in the interstellar medium.  This implies that the dust in the cluster seems to be relaxed with reference to the interstellar magnetic field i.e., the dust grains in the cluster are magnetically aligned with the large-scale magnetic field at the location of the cluster.

\subsection{Dust distribution towards Czernik 3}\label{sec:4.3}
\begin{figure}[!h]
	\plotone{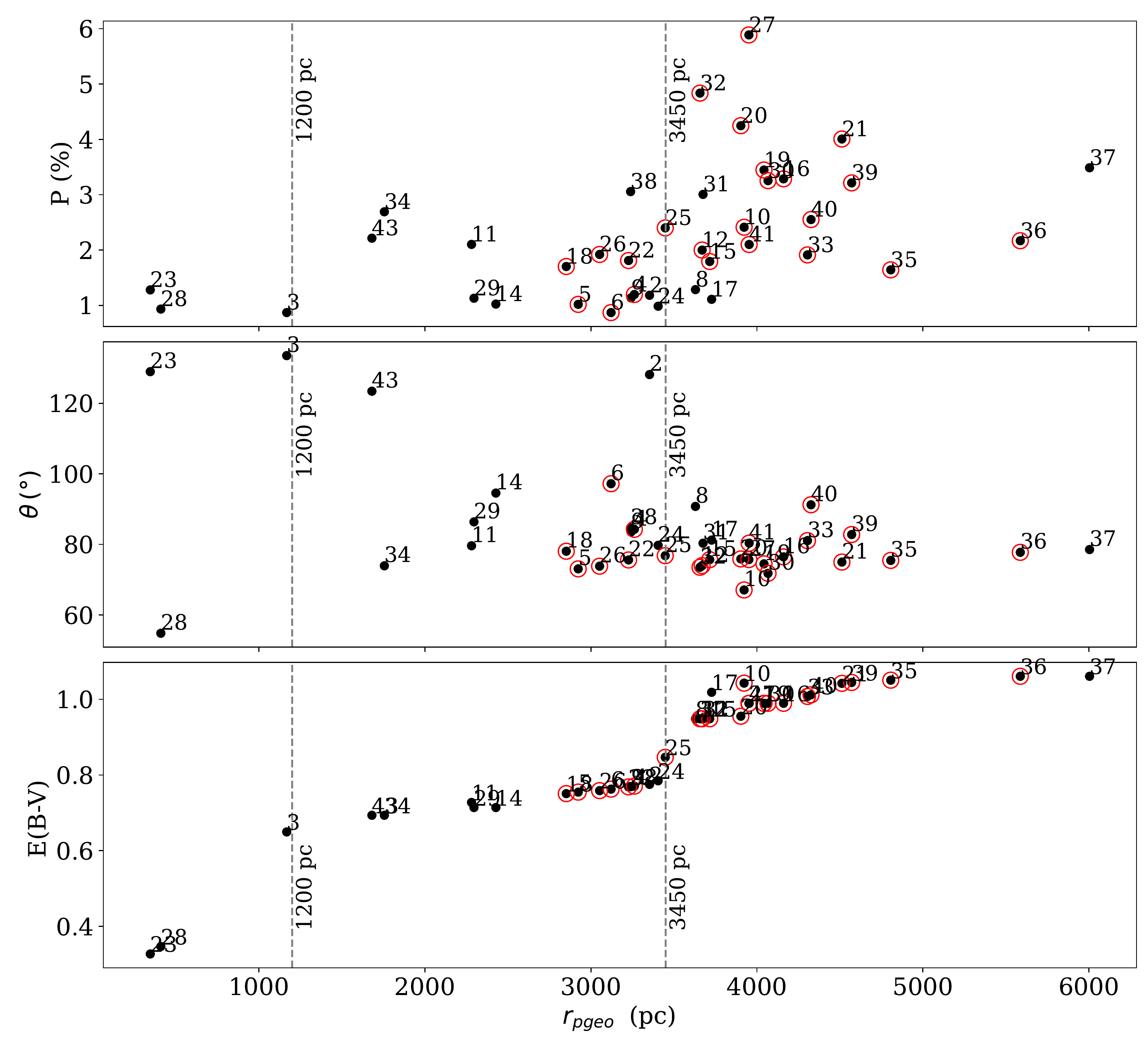}
    \caption{Variation of degree of polarization, polarization angle of stars observed in the line of sight of Czernik 3} and $50^{th}$ percentile reddening (E(B-V)) as a function of distance.
    \label{fig:percent}
\end{figure}

Polarization value shows a change when light from the stars encounters dust layers at different distances in the line of sight. The degree of polarization increases after passing through each dust layer if the orientation of the magnetic field is uniform in all dust clouds and becomes depolarized if the orientation of the magnetic field are different. The number of such changes in the degree of polarization as well as in extinction corresponds to the number of dust layers. 
Thus, the distribution of dust along the line of sight can be studied by analysing polarization with distance.  
 In order to quantify the distances of the dust clouds, we have used distance information ($r_{pgeo}$) of individual stars from \citet{bailer2021}. One star, \#20, does not have the $r_{pgeo}$ value. Consequently, we have used the $r_{geo}$ (distance estimated from EDR3 parallax using Geometric prior only) value for that source.  Figure \ref{fig:percent} shows the variation of polarization, position angle, and E(B-V) with distance ($r_{pgeo}$).  The E(B-V) values are taken from \citet{Green2019} \citep[calculated using a python based script presented by][]{dustPython}.  The cluster members are marked as red open circles. Middle panel of Figure \ref{fig:percent} reveals a significant change in the polarization angle between star \#43 and \#34.  This indicates the presence of a dust layer at a distance $<$ 1800 pc (upper limit).  We also note that there is an increase (top panel) in the degree of polarization just beyond $\sim 1000$ pc.  Due to the lack of polarization data for stars below 2000 pc, it is difficult to constrain the distance of the foreground dust cloud. 
 However, the E(B-V) values  shows a jump at a distance $<$ 1200 pc (see bottom panel of Figure \ref{fig:percent}). This could be because of the presence of LDN1306, a Lynd's dark cloud \citep{Lynds1962}. The cloud has an angular extent of $\sim 200^\prime$ \citep{Dutra2002} and it is present south-east to the cluster with angular separation of $\sim 31^\prime$. The reported $50^{th}$ percentile distance to the cloud is $941$ pc \citep{darkcloudDist2020}, which is consistent with the location of the observed change in the degree of polarization as well as in E(B-V) close to $\sim 1200$ pc. 
Stars beyond 2000 pc have large uncertainties (see Figures \ref{fig:PolDist} (a) and (b), along with Figure \ref{fig:percent}) associated with the distance ($r_{pgeo}$). However, E(B-V) value shows a systematic change with distance, indicating another dust layer around 3450 pc. The spread in distance of the member stars (shown as red open circles in Figure \ref{fig:percent}) is the result of large fractional parallax errors of these distant sources. The presence of a dust layer around 3450 pc nearly coinciding with the average distance to the cluster ($3600 \pm 800$ pc) indicates the following possibilities:
\begin{enumerate}
    \item the cluster may be embedded in the dust layer
    \item the cluster is passing through a dense region of the galactic plane and coincidentally has a similar distance as that of a cloud near that location
    \item the dust layer is present just before the cluster.
\end{enumerate}
To confirm the presence of the second dust layer, we consider the fact that dust and HI are correlated in the diffuse ISM \citep[e.g.,][]{Bohlin}.  Therefore, we use kinematic information of HI line emission spectra from HI4PI survey \citep{HI4PI} to infer the distribution of atomic gas along the line of sight in the selected region. The spectrum (blue line in Figure \ref{fig:HICO}) reveals the existence of four well separated velocity peaks (-114.7, -100.4, -43.8, -13.0 K~ms$^{-1}$) and implies the neutral ISM mass is distributed in at least 4 spatially distinct components along the line of sight. The calculated kinematic distance of corresponding peak is given in Table \ref{tab:kinematic} (detailed information is given in Appendix \ref{Ap2}).
Clouds with peak velocity -114.7 km~s$^{-1}$ and -100 km~s$^{-1}$ are in the  background of the cluster (distance $>$ 6 kpc). Thus, they will not contribute to the observed polarization. The main contribution comes from the clouds having distance of $3.17^{+0.63}_{-0.59}$ kpc and $0.74^{+0.52}_{-0.52}$ which is consistent with the color excess and polarization jumps observed in the Figure \ref{fig:percent}. The same two clouds with radial velocity of  $-43.8$ km~s$^{-1}$, $-13.0$ km~s$^{-1}$ are also seen in the molecular data (orange line in Figure \ref{fig:HICO}), i.e., spectral information of $^{12}$CO from \citet{12CO}. This confirms  the presence of at-least two clouds along our line of sight.

In addition, the variation in E(B-V) and degree of polarization of the member stars with distance (Figure \ref{fig:percent}) also suggests patchy extinction present towards the Czernik 3 cluster. 
\subsection{General trend of dust distribution: a signature of Inter-arm region}\label{sec:4.4}
\begin{figure}[!h]
	\plotone{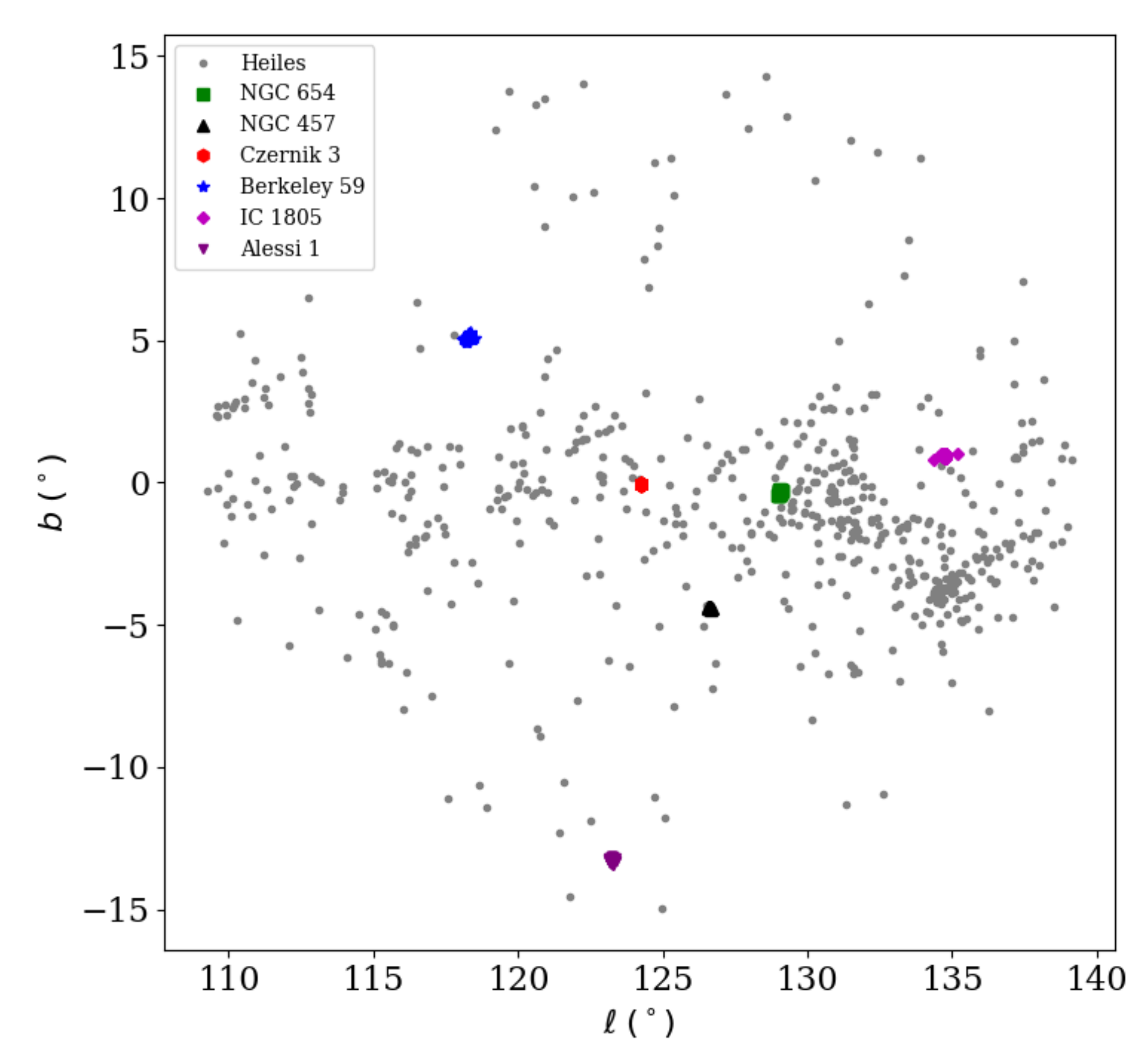}
    \caption{Distribution of clusters (color $\&$ symbols represents different cluster) and \citet{Heiles2000} stars (gray dots) in Galactic coordinates ($\ell$-$b$) centered on Czernik 3.}
    \label{fig:lb}
\end{figure}

In this section, we study the global properties of dust distribution that can be derived by combining the polarization data of Czernik 3, one of the most distant clusters with polarization data, and other clusters present within 15$^\circ$ region around it (see Table \ref{tab:clusters}) along with the \citet{Heiles2000} stars (Figure \ref{fig:lb}).

\begin{figure}[!h]
	\plotone{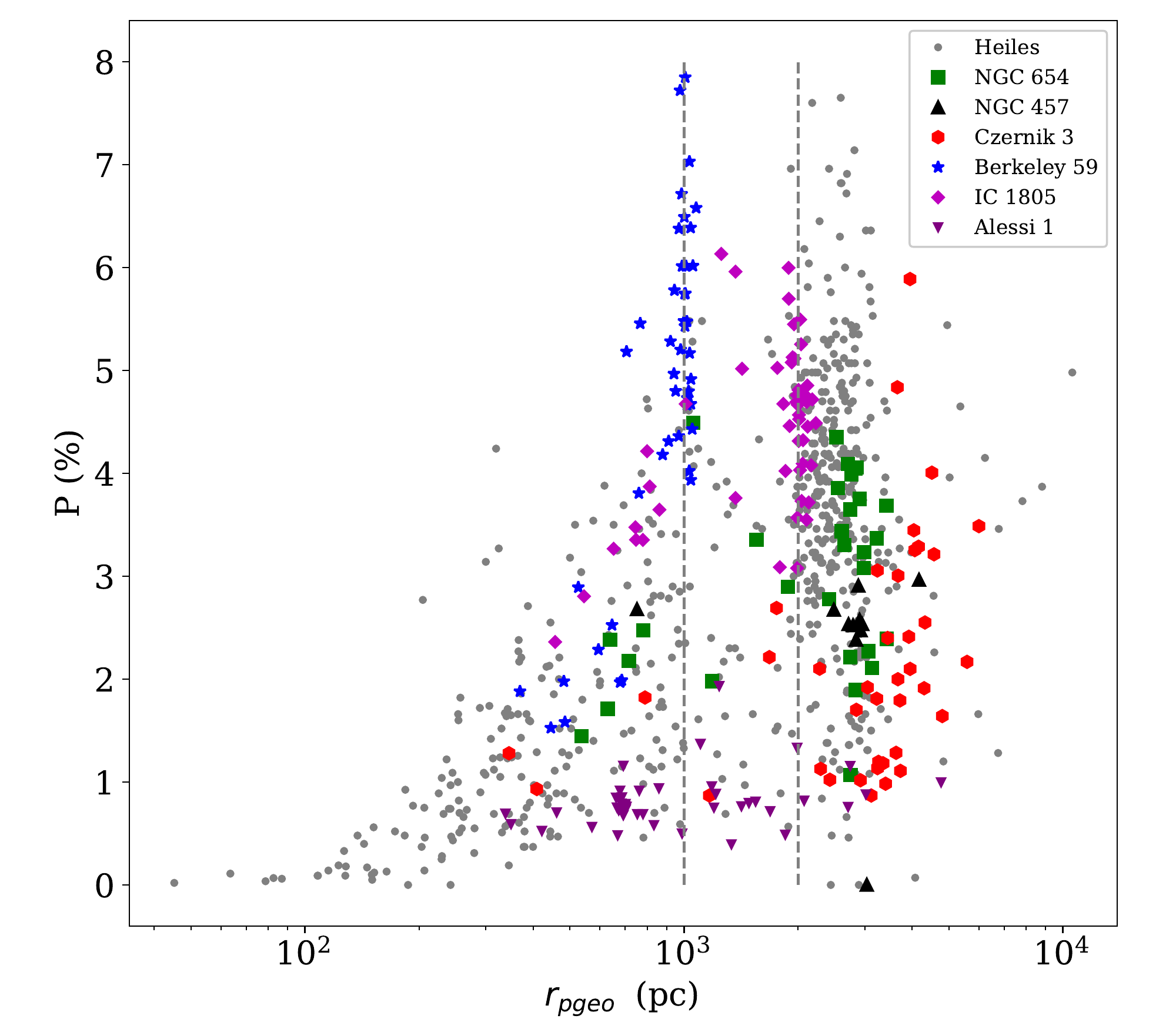}
    \caption{Variation of degree of polarization of stars towards different clusters as a function of distance. \citep{Heiles2000} polarization within 15$^\circ$ is also added in gray colored dots.}
    \label{fig:general}
\end{figure}

The polarization measurements of these clusters were converted to equivalent values in the \textit{Sloan - i} band by utilizing the respective $\lambda_{max}$ and $P_{max}$ values in the Serkowski law for wavelength dependence of polarization.
The degree of polarization of all the available clusters in \textit{sloan i}-band, along this direction is plotted as a function of distance ($r_{pgeo}$, pc) in Figure \ref{fig:general}.
The polarization of stars within $15^\circ$ of Czernik 3 from  \citet{Heiles2000} is also added in gray color. It is seen that the degree of polarization gradually increases with distance as expected from the uniform dust distribution close to the solar neighbourhood (in the local arm) but shows a sudden increase due to a dust patch at $\sim$700 pc \citep{Bk59}.  There is a drop in the  number of stars between 1-2 kpc in all the cluster fields 
as well as the field stars from the Heiles catalog and there is no systematic change in polarization in this region.
The polarization angle also remains consistent in the two regions ($<$ 1kpc \& $>$ 2kpc). 
This peculiar decrease in number of stars and no observed systematic change in polarization indicates that this region may contain minimal dust to affect the polarization of the background stars.  This decrease in the number of stars with polarization measurement is not due to sample incompleteness.  Rather, we note that the overall stellar number density (detection in the 2MASS bands with quality flag `AAA') also shows a significant drop between 1 and 2 Kpc. The decline in stellar density as well as dust density (from polarization data) suggests that the 1-2 Kpc region could be the inter-arm region between the local arm (below 1 kpc) and the Perseus arm \citep[$>$ 2kpc, ][]{perseusDist}. 

\begin{figure}[!ht]
	\plotone{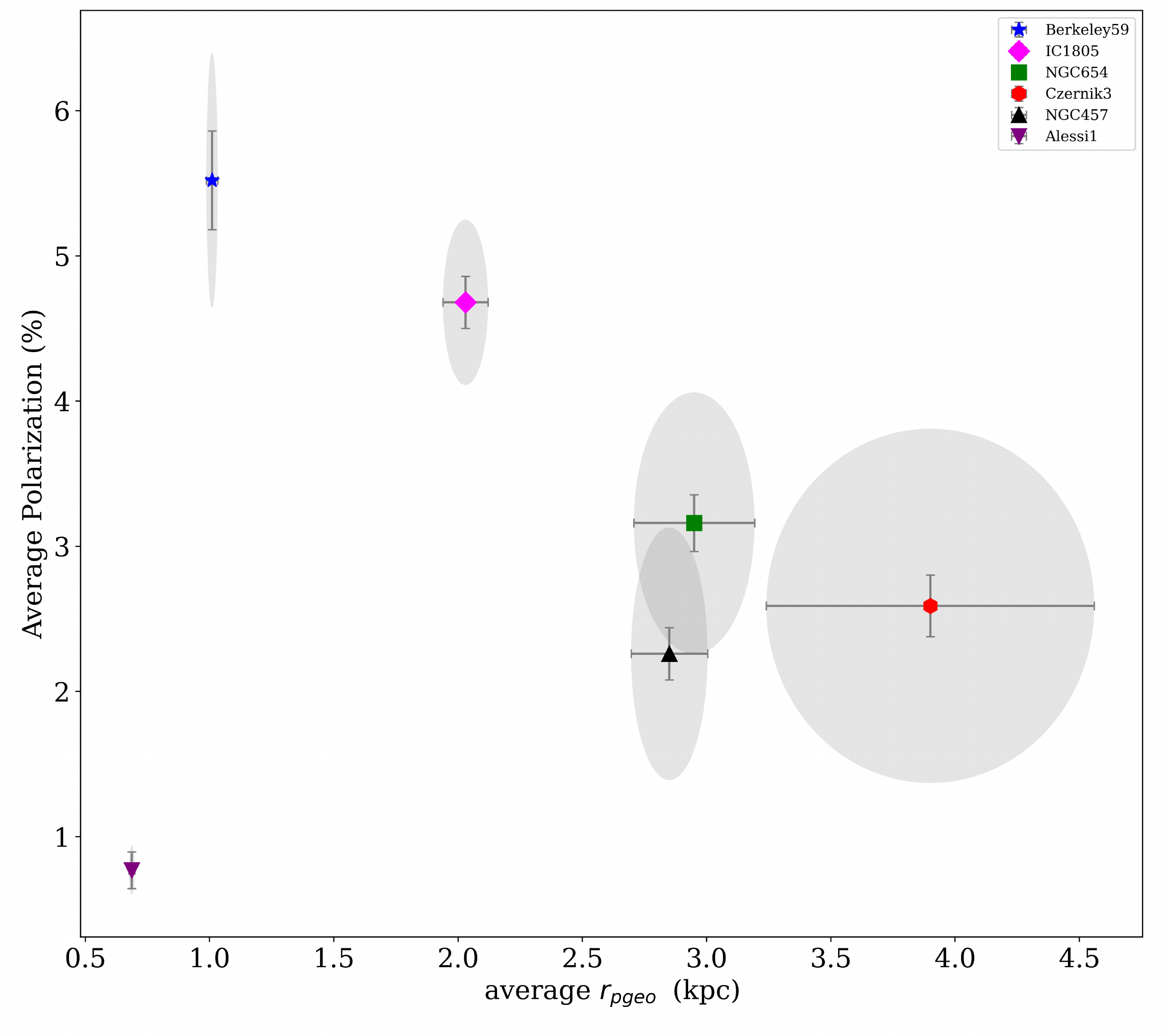}
    \caption{Variation of average polarization towards each cluster as a function of mean distance to the corresponding cluster.}
    \label{fig:clusters}
\end{figure}

The variation of average degree of polarization of each cluster with the mean distance of the member stars (membership probability $> 90\%$) with polarization measurement is depicted in Figure \ref{fig:clusters}. 
This figure shows a decrease in mean polarization of the cluster with distance. 

We note that Alessi 1 is a high latitude cluster (see Figure \ref{fig:lb}), suffering lesser dust extinction as compared to other clusters, hence showing exceptionally low degree of polarization in comparison to the others. On the other hand Berkeley 59 is a young cluster having intra-cluster 
dust due to which many cluster members have intrinsic (non-interstellar) polarization up to 2\% \citep{Bk59}.  The other clusters considered here, are relatively older.  With these caveats in mind, we still see an overall decrease in the degree of polarization with distance (Figure \ref{fig:clusters}), which could be due to decrease in polarization efficiency with increasing path length. 
\section{Conclusions}\label{sec:5}
In this work, we presented the linear polarization study towards a distant open cluster, Czernik 3 and 15$^\circ$ region around it, with the aim to estimate the dust distribution on different spatial  scales. We also redefined the membership of Czernik 3 as well as the other nearby clusters, having polarization information, considered in this work. The polarimetric observations of Czernik 3 were carried out using the EMPOL instrument at the 1.2 m telescope at Mount Abu. A total of 42 stars are observed towards the core region of the cluster in \textit{Sloan i}-band. We have arrived at the following conclusions from our study:
\begin{enumerate}
\item From the observed changes in the degree of polarization and extinction, we infer the presence of two dust clouds along the line of sight towards Czernik 3 cluster, at distances of $\sim 1$ kpc and $\sim 3.4$ kpc. The first cloud corresponds to the Lynds dark nebula LDN 1306. 
\item The large range in the observed polarization  ($0.87\% \le P \le 5.89\%$) of Czernik 3 stars may result from a non-uniform polarization efficiency. It could also arise from the non-uniform/patchy distribution of dust towards the cluster. We observed more dust density in the core of the cluster as compared to the peripheral regions.
\item The cluster membership of Czernik 3 as well as other nearby clusters is redefined using the astrometric and photometric data from the latest Gaia EDR3.  
The distance to the Czernik 3 cluster is well constrained using member stars present in the core and it comes out to be $3.6 \pm 0.8$ kpc. 
\item The polarization efficiency for most of the observed stars towards Czernik 3 is less than the upper limit of the empirical relation for ISM polarization. 
Two stars with exceptionally high polarization are closer to the center of the cluster indicating different  polarization  behaviour as compared to the general diffuse ISM.  They may also be intrinsically polarized.
\item For the clusters studied here, we find that the average polarization of the cluster decreases with distance to the cluster, indicating a decrease in polarization efficiency with increasing path length.
\item There is a decrease in dust as well as stellar density in the distance range of 1 to 2 kpc in a 15$^\circ$ wide region around Czernik 3.
This is speculated to be the inter-arm region between the local arm and the Perseus arm.

\end{enumerate}
\section{Acknowledgement}
We thank the referee, Dr Enrique Lopez-Rodriguez, for his careful evaluation of the manuscript and the insightful comments which helped to improve the paper.  
We thank the local staff at Mount Abu Observatory, PRL for their help during observations.  We would also like to thank Prof U C Joshi, Prof K S Baliyan and our colleagues in the Astronomy and Astrophysics division, PRL for useful discussions. 
Work at Physical Research Laboratory is supported by the Department of Space, Govt. of India. 
A part of this work has made use of data from the European Space Agency (ESA) mission Gaia\footnote{https://www.cosmos.esa.int/web/gaia}, processed by the Gaia Data Processing and Analysis Consortium (DPAC\footnote{https://www.cosmos.esa.int/web/gaia/dpac/consortium}). Funding for the DPAC is provided by national institutions, in particular, the institutions participating in the Gaia Multilateral Agreement. This research made use of Astropy,\footnote{http://www.astropy.org} a community-developed core Python package for Astronomy \citep{astropyI, astropyII}. We have also used the VizieR catalogue access tool, CDS, Strasbourg, France. 
\appendix
\section{Cluster membership through photometric and polarimetric approach}\label{ap:1}
It is expected that all the member stars are behind the same dust clouds, thus the colour-excess of the member stars should be comparable  to the mean colour-excess of the cluster.  In comparison, the field stars suffer varying extinction based on whether they are present foreground or background to the cluster.  \citet{membership} have shown that the TCD together with the polarization can be used to distinguish the cluster members from the field stars.

\begin{figure*}
   \gridline{\fig{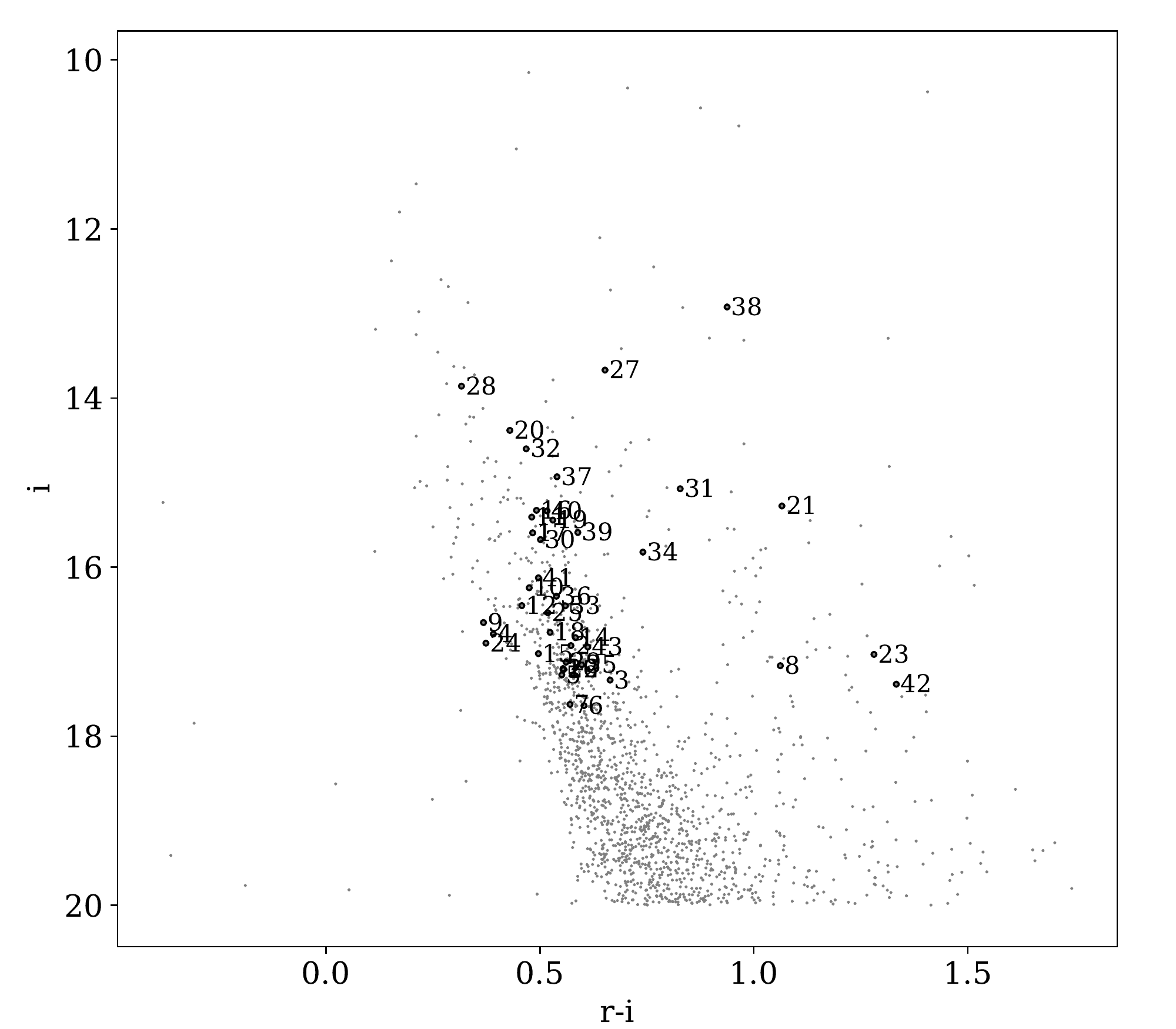}{0.5\textwidth}{(a)}
            \fig{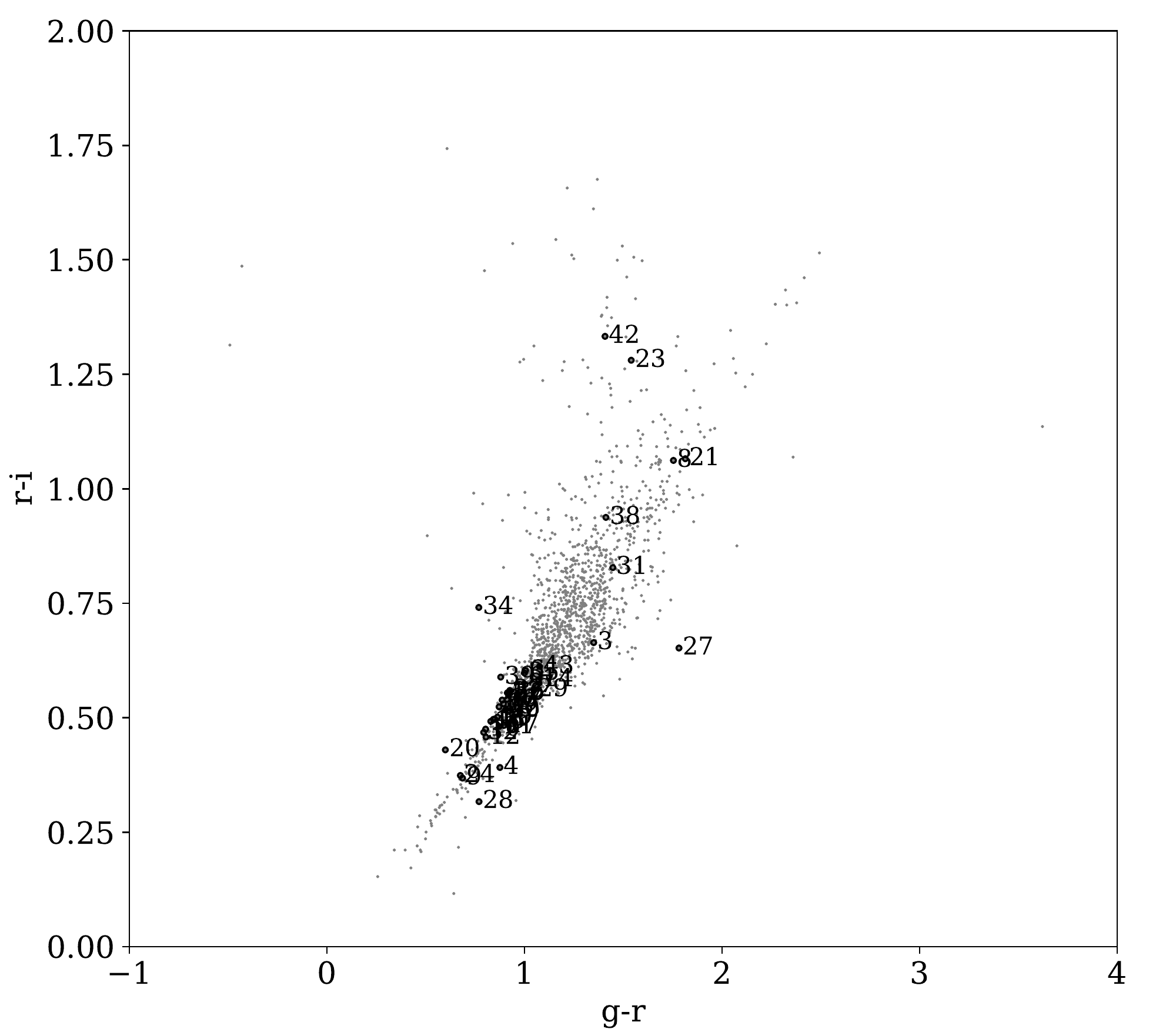}{0.5\textwidth}{(b)}
    }
    \caption{Color magnitude  (r-i versus i (a)) and color-color (g-r versus r-i in (b)) using PanSTARRS data. The small dots (gray color) represents the $10^\prime \times 10^\prime$ field towards the Czernik 3 cluster. Black points with labels are the stars having polarization observations.}
    \label{fig:newcmd}
\end{figure*}

We now consider the stars in the $q-u$ plane of Figure \ref{fig:stokes} in the context of their PanSTARRS magnitudes and colours.  The PanSTARRS apparent color-magnitude diagram and TCD for a $10^\prime \times 10^\prime$ region around the cluster is shown in Figure \ref{fig:newcmd} (a) and (b). The ID of the stars listed in the Table \ref{tab:obs} are labeled in these two figures.

The group-1 stars(\#2, \#3, \#23, and \#43) of Figure \ref{fig:pol_angle}  have polarization angle larger than that of the Galactic plane and lie outside the mean $1\sigma$ box on $qu$ plane (Figure \ref{fig:stokes}) with range of $ -0.8 < q< -0.4$ and $-2.0 < u < -0.9$. Stars \#3 and \#23 show more reddening (see Figure \ref{fig:newcmd} (a) and (b)) than the bulk of the stars. Hence, these two stars are not member stars according to both approaches (photometric and polarimetric). Though the color of stars \#2 and \#43 are similar to the member stars, their location in the $qu$ plot and their position angle values indicate deviation from polarization angle of member stars. Hence, these stars are also considered as non-members.

Star \#28 have q $>$ -0.5 (outside mean $1\sigma$ box in Figure \ref{fig:stokes}) just like \#2, \#3 and \#23 and have large errors associated with their polarization values. It has a low value of polarization - $0.93\% \pm 0.23\%$ and polarization angle $54.8 \pm 6.8$ which is quite different from the rest of the stars.  This 
indicates that \#28 could be a foreground star, which is consistent with relatively bluer color r-i = 0.31 and a Gaia EDR3 distance of 409 pc.  

Stars \#27 and \#38 are reddened in colour (see Figure \ref{fig:newcmd} (a) and (b)). Star \#27 shows highest polarization ($5.89\% \pm 0.15\%$) in our data set while the polarization angle ($75^\circ.7 \pm 0^\circ.7$) is similar to the member stars (group 3 stars). It may be a member star having some intrinsic component of polarization. On the other hand, Star \#38 is outside the mean $1\sigma$ boundary of group 3 in $q-u$ plane (see Figure \ref{fig:stokes}) and shows a deviation in polarization angle ($84^\circ.7 \pm 1^\circ.5$) from mean polarization angle of group 3 stars. It is highly probable that this star may also have intrinsic polarization.  Discussion about its cluster membership is deferred to the  Section \ref{astrometry}. The rest of the distribution in the $q-u$ plane is scattered, however, the region 3  of Figure \ref{fig:stokes} contains the most probable cluster members. 

Considering only photo-polarimetric approach, the contribution to the cluster membership, by the field stars, can be overestimated if there is a common dust layer foreground to the field stars and the cluster members. Hence, polarization alone cannot be used to estimate the membership probability of distant open clusters.
\section{Variation of polarization with distance}
\begin{figure*}
   \gridline{\fig{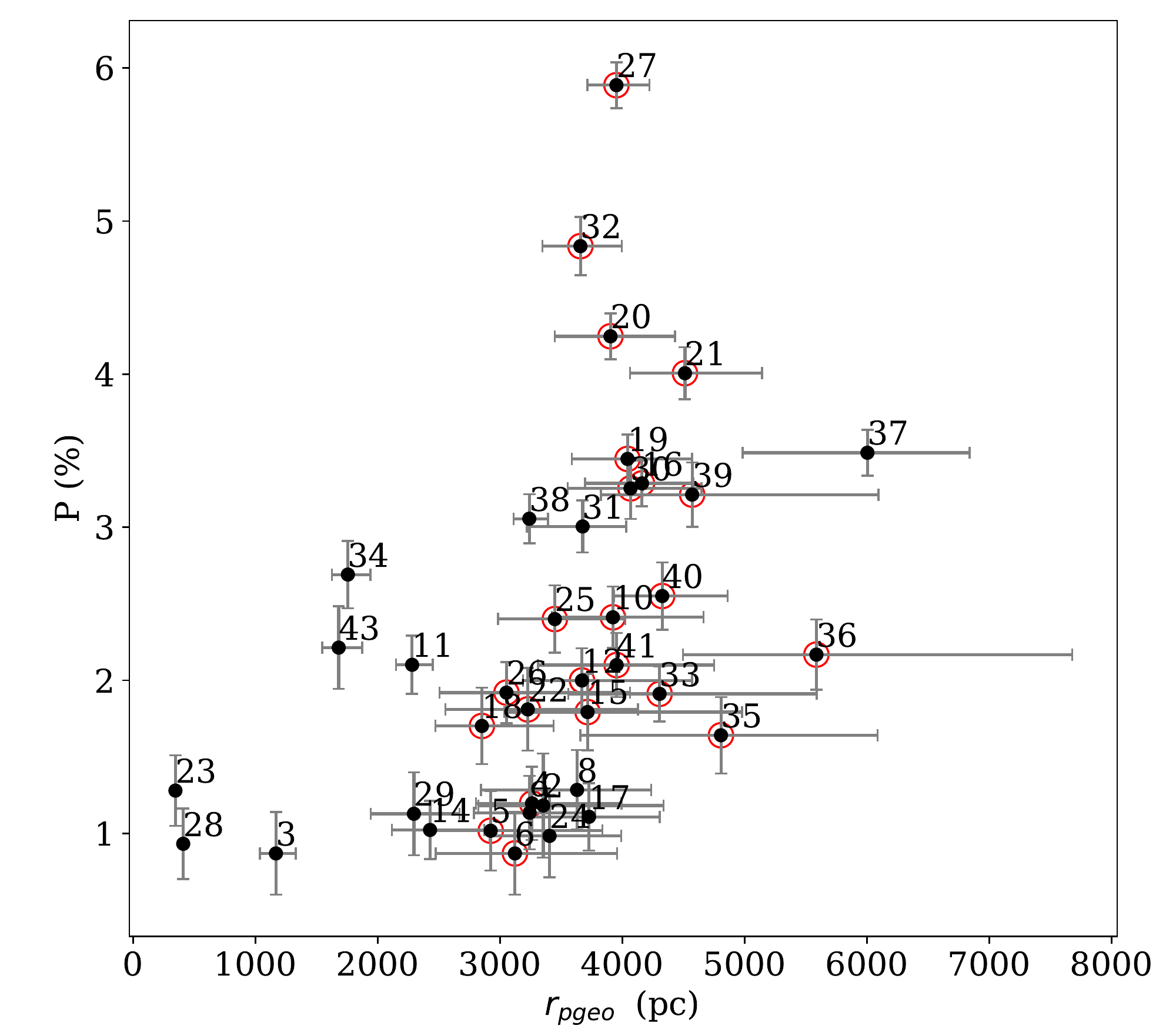}{0.5\textwidth}{(a) }
            \fig{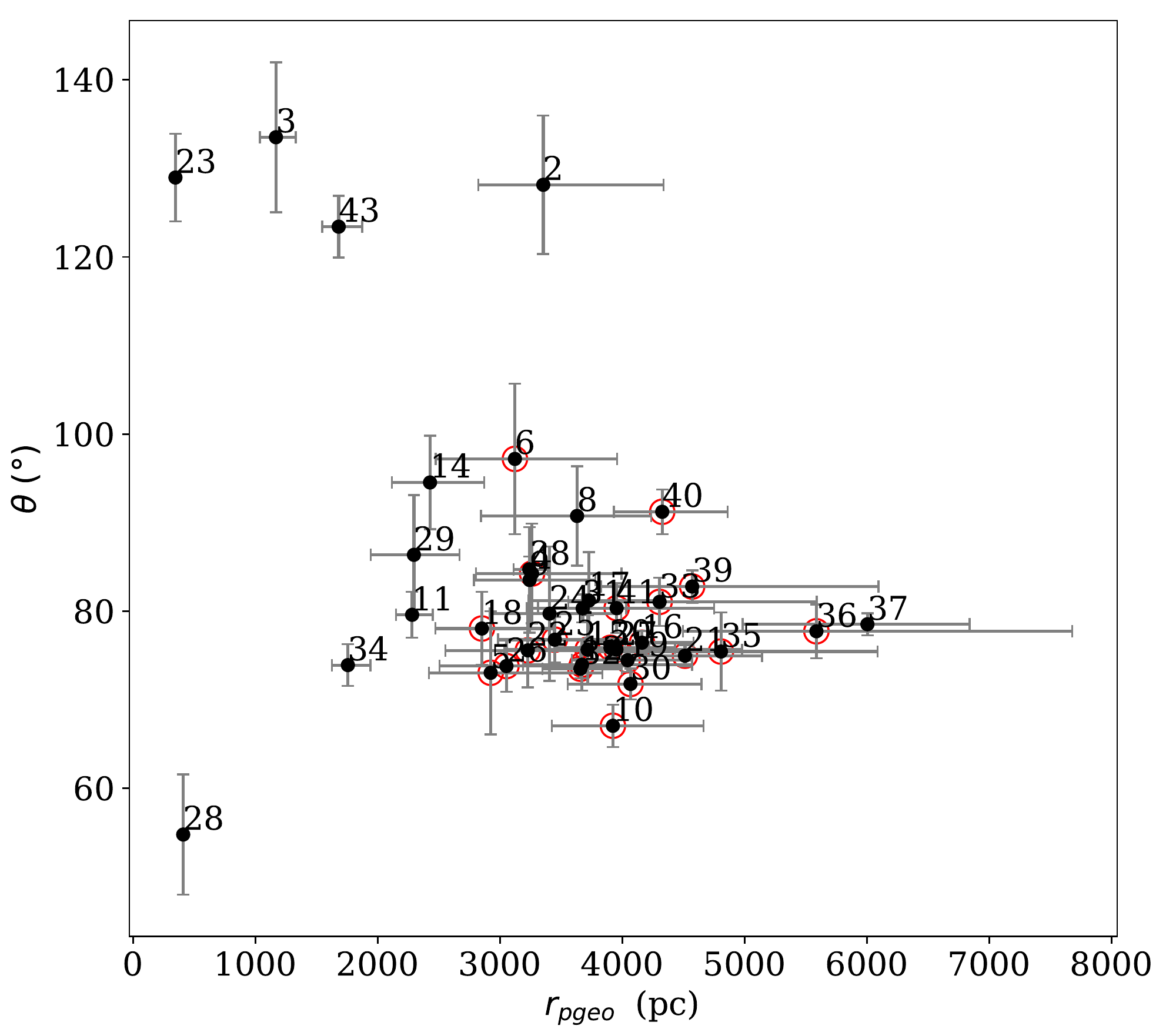}{0.5\textwidth}{(b)}
    }
    \caption{Variation of degree of polarization (a) and polarization angle (in b) with distance of observed stars in Czernik 3 direction. The errors in the Gaia EDR3 distances ($r_{pgeo}$) and degree of polarization/polarization angle are also shown. The member stars are marked as red open circles.}
    \label{fig:PolDist}
\end{figure*}

The variation of degree of polarization and polarization angles as a function of distance ($r_{pgeo}$) along with their corresponding error-bars are displayed in Figure \ref{fig:PolDist} (a) and (b) respectively. Detailed discussion is available in section \ref{sec:4.3}

\section{Distance calculation of H1 and $^{12}$CO clouds}\label{Ap2}
\begin{figure}[!h]
	\plotone{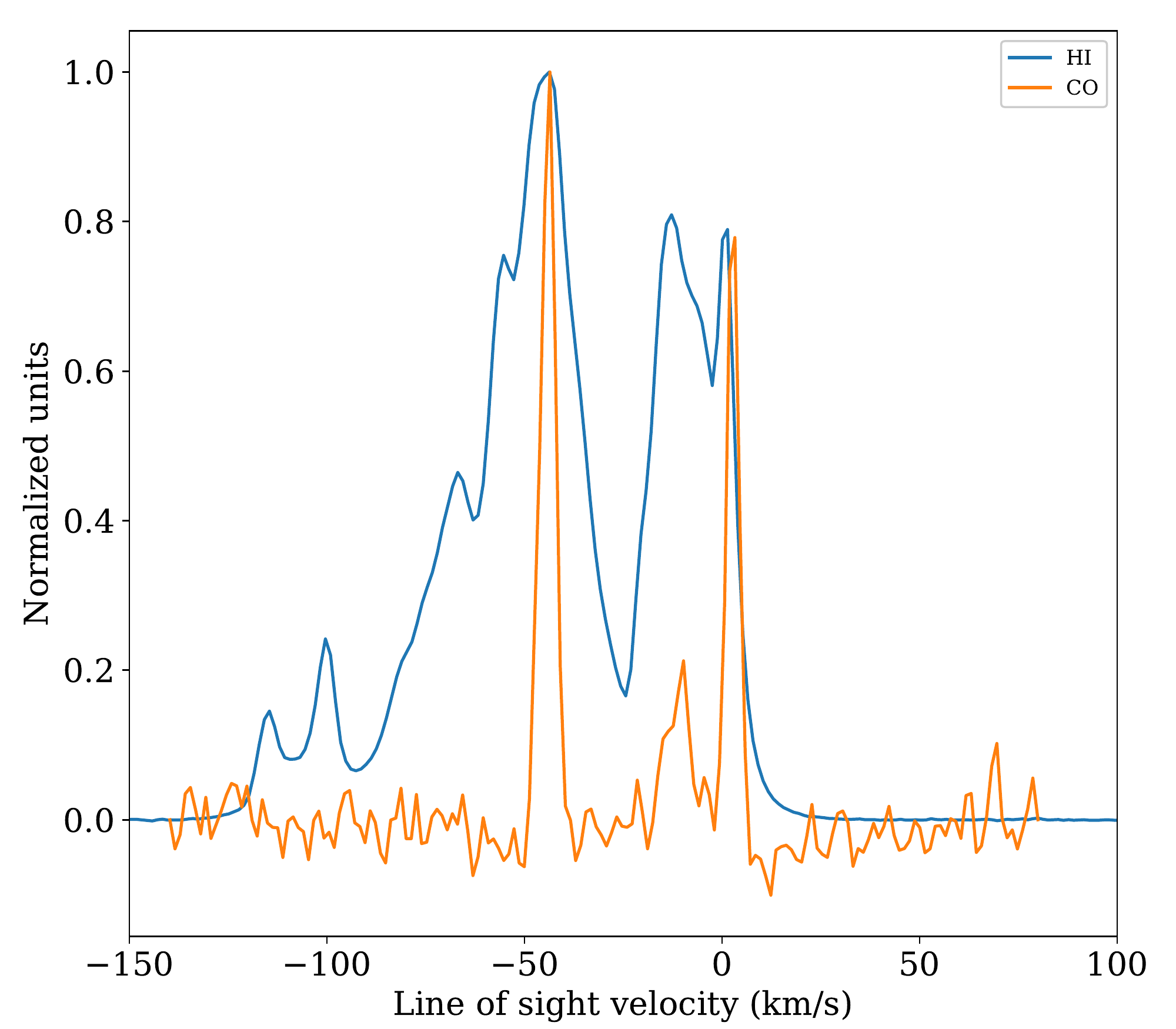}
    \caption{Spectrum of neutral hydrogen clouds $HI$- 21 cm line (in blue color) within $16^\prime$ of the center of Czernik 3 cluster and $^{12}$CO (orange) within $30^\prime$ of the center of the cluster. Y-axis represents the normalized brightness temperature for HI and normalized antenna temperature for $^{12}$CO.}
    \label{fig:HICO}
\end{figure}

The spectral information of 21-cm neutral hydrogen line (HI) within $16^\prime$  region of Czernik 3 cluster centered at $\alpha_{J2000} = 01^h03^m06^s.9; \delta_{J2000}= 62^\circ47^\prime00^{\prime\prime}$ is taken from the HI4PI survey (angular resolution: 16$^{\prime}$.2) and is shown as blue curve in Figure \ref{fig:HICO}. The HI peaks show the presence of multiple clouds in the line of sight with peak velocity of $\sim0$ km~s$^{-1}$, $\sim-13.0$ km~s$^{-1}$, $\sim -43 $ km~s$^{-1}$, $\sim  -100.4$ km~s$^{-1}$, and $\sim -114.7$ km~s$^{-1}$. There could be two more clouds with peak velocities of $\sim -55.3$ and $\sim -67$ km~s$^{-1}$.
The orange curve in the Figure \ref{fig:HICO} represents the  spectral information of $^{12}$CO (angular resolution: $\sim 30^{\prime}$) within the $30 ^\prime$ region around the center of the cluster and indicates three clouds along the line of sight with peak velocity closer to the first three HI clouds.  
The kinematic distance of the HI and $^{12}$CO  clouds are calculated from the web interface\footnote{\url{http://bessel.vlbi-astrometry.org/revised_kd_2014?}} corresponding to the source code provided by \citet{Reid2009KinematicDistance}. We have used the Galactic parameters :
$\Theta_{0} = 240$ km~s$^{-1}$, $R_{0}=8.34$ kpc, $\frac{d\Theta}{dR}=-0.2$ km~s$^{-1}$
kpc$^{-1}$, $U_{\odot}= 10.7$ km~s$^{-1}$, $V_{\odot}=15.6$ km~s$^{-1}$, $W_{\odot} =  8.9$ km~s$^{-1}$, $\bar{U_{s}}=2.9$ km~s$^{-1}$, $\overline{V_{s}}=-1.6$ km~s$^{-1}$ following A5 model of \citet{Reid2014parameters}. The corresponding distances are tabulated in Table \ref{tab:kinematic}.

Three clouds are seen in common (at similar velocities) in both HI as well as $^{12}$CO. The neutral hydrogen, HI traces the diffuse medium while $^{12}$CO traces the outer regions of dense molecular clouds. Therefore, we consider clouds having $^{12}$CO data coinciding with HI radial velocity peaks around $\sim -13$ km~s$^{-1}$ and $\sim -43.7$ km~s$^{-1}$.

\begin{deluxetable*}{cc}
\tablecaption{Peak velocity in HI spectrum in $5^\prime$ field centered on cluster center and corresponding kinematic distance in kpc. \label{tab:kinematic} }
  
\tablewidth{0pt}
\tablehead{
\colhead{Line of sight velocity (Km~s$^{-1}$)} & \colhead{Kinematic distance (kpc) }
}
\startdata
-114.7 & $12.86^{+1.74}_{-1.50}$\\
-100.4 & $10.0^{+1.30}_{-1.15}$\\
-67.0 & $5.43^{+0.79}_{-0.65}$\\
-55.3 & $4.23^{+0.70}_{-0.65}$\\
-43.7 &  $3.17^{+0.63}_{-0.59}$\\
-13.0 & $0.74^{+0.52}_{-0.52}$\\
\enddata
\end{deluxetable*}

\bibliography{ref}{}

\begin{thebibliography}{}
\expandafter\ifx\csname natexlab\endcsname\relax\def\natexlab#1{#1}\fi
\providecommand{\url}[1]{\href{#1}{#1}}
\providecommand{\dodoi}[1]{doi:~\href{http://doi.org/#1}{\nolinkurl{#1}}}
\providecommand{\doeprint}[1]{\href{http://ascl.net/#1}{\nolinkurl{http://ascl.net/#1}}}
\providecommand{\doarXiv}[1]{\href{https://arxiv.org/abs/#1}{\nolinkurl{https://arxiv.org/abs/#1}}}

\bibitem[{{Astropy Collaboration} {et~al.}(2013){Astropy Collaboration},
  {Robitaille}, {Tollerud}, {Greenfield}, {Droettboom}, {Bray}, {Aldcroft},
  {Davis}, {Ginsburg}, {Price-Whelan}, {Kerzendorf}, {Conley}, {Crighton},
  {Barbary}, {Muna}, {Ferguson}, {Grollier}, {Parikh}, {Nair}, {Unther},
  {Deil}, {Woillez}, {Conseil}, {Kramer}, {Turner}, {Singer}, {Fox}, {Weaver},
  {Zabalza}, {Edwards}, {Azalee Bostroem}, {Burke}, {Casey}, {Crawford},
  {Dencheva}, {Ely}, {Jenness}, {Labrie}, {Lim}, {Pierfederici}, {Pontzen},
  {Ptak}, {Refsdal}, {Servillat}, \& {Streicher}}]{astropyI}
{Astropy Collaboration}, {Robitaille}, T.~P., {Tollerud}, E.~J., {et~al.} 2013,
  \aap, 558, A33, \dodoi{10.1051/0004-6361/201322068}

\bibitem[{{Astropy Collaboration} {et~al.}(2018){Astropy Collaboration},
  {Price-Whelan}, {Sip{\H{o}}cz}, {G{\"u}nther}, {Lim}, {Crawford}, {Conseil},
  {Shupe}, {Craig}, {Dencheva}, {Ginsburg}, {VanderPlas}, {Bradley},
  {P{\'e}rez-Su{\'a}rez}, {de Val-Borro}, {Aldcroft}, {Cruz}, {Robitaille},
  {Tollerud}, {Ardelean}, {Babej}, {Bach}, {Bachetti}, {Bakanov}, {Bamford},
  {Barentsen}, {Barmby}, {Baumbach}, {Berry}, {Biscani}, {Boquien}, {Bostroem},
  {Bouma}, {Brammer}, {Bray}, {Breytenbach}, {Buddelmeijer}, {Burke},
  {Calderone}, {Cano Rodr{\'\i}guez}, {Cara}, {Cardoso}, {Cheedella}, {Copin},
  {Corrales}, {Crichton}, {D'Avella}, {Deil}, {Depagne}, {Dietrich}, {Donath},
  {Droettboom}, {Earl}, {Erben}, {Fabbro}, {Ferreira}, {Finethy}, {Fox},
  {Garrison}, {Gibbons}, {Goldstein}, {Gommers}, {Greco}, {Greenfield},
  {Groener}, {Grollier}, {Hagen}, {Hirst}, {Homeier}, {Horton}, {Hosseinzadeh},
  {Hu}, {Hunkeler}, {Ivezi{\'c}}, {Jain}, {Jenness}, {Kanarek}, {Kendrew},
  {Kern}, {Kerzendorf}, {Khvalko}, {King}, {Kirkby}, {Kulkarni}, {Kumar},
  {Lee}, {Lenz}, {Littlefair}, {Ma}, {Macleod}, {Mastropietro}, {McCully},
  {Montagnac}, {Morris}, {Mueller}, {Mumford}, {Muna}, {Murphy}, {Nelson},
  {Nguyen}, {Ninan}, {N{\"o}the}, {Ogaz}, {Oh}, {Parejko}, {Parley}, {Pascual},
  {Patil}, {Patil}, {Plunkett}, {Prochaska}, {Rastogi}, {Reddy Janga},
  {Sabater}, {Sakurikar}, {Seifert}, {Sherbert}, {Sherwood-Taylor}, {Shih},
  {Sick}, {Silbiger}, {Singanamalla}, {Singer}, {Sladen}, {Sooley},
  {Sornarajah}, {Streicher}, {Teuben}, {Thomas}, {Tremblay}, {Turner},
  {Terr{\'o}n}, {van Kerkwijk}, {de la Vega}, {Watkins}, {Weaver}, {Whitmore},
  {Woillez}, {Zabalza}, \& {Astropy Contributors}}]{astropyII}
{Astropy Collaboration}, {Price-Whelan}, A.~M., {Sip{\H{o}}cz}, B.~M., {et~al.}
  2018, \aj, 156, 123, \dodoi{10.3847/1538-3881/aabc4f}

\bibitem[{{Axon} \& {Ellis}(1976)}]{axon1976catalogue}
{Axon}, D.~J., \& {Ellis}, R.~S. 1976, \mnras, 177, 499,
  \dodoi{10.1093/mnras/177.3.499}

\bibitem[{{Bailer-Jones} {et~al.}(2021){Bailer-Jones}, {Rybizki}, {Fouesneau},
  {Demleitner}, \& {Andrae}}]{bailer2021}
{Bailer-Jones}, C.~A.~L., {Rybizki}, J., {Fouesneau}, M., {Demleitner}, M., \&
  {Andrae}, R. 2021, \aj, 161, 147, \dodoi{10.3847/1538-3881/abd806}

\bibitem[{{Balaguer-N{\'u}nez} {et~al.}(1998){Balaguer-N{\'u}nez}, {Tian}, \&
  {Zhao}}]{BN98}
{Balaguer-N{\'u}nez}, L., {Tian}, K.~P., \& {Zhao}, J.~L. 1998, \aaps, 133,
  387, \dodoi{10.1051/aas:1998324}

\bibitem[{{Bellini} {et~al.}(2009){Bellini}, {Piotto}, {Bedin}, {Anderson},
  {Platais}, {Momany}, {Moretti}, {Milone}, \& {Ortolani}}]{bellini2009}
{Bellini}, A., {Piotto}, G., {Bedin}, L.~R., {et~al.} 2009, \aap, 493, 959,
  \dodoi{10.1051/0004-6361:200810880}

\bibitem[{{Berdyugin} {et~al.}(2014){Berdyugin}, {Piirola}, \&
  {Teerikorpi}}]{largescale}
{Berdyugin}, A., {Piirola}, V., \& {Teerikorpi}, P. 2014, \aap, 561, A24,
  \dodoi{10.1051/0004-6361/201322604}

\bibitem[{{Bertin} \& {Arnouts}(1996)}]{sextractor}
{Bertin}, E., \& {Arnouts}, S. 1996, \aaps, 117, 393,
  \dodoi{10.1051/aas:1996164}

\bibitem[{{Bisht} {et~al.}(2017){Bisht}, {Yadav}, \& {Durgapal}}]{Bisht}
{Bisht}, D., {Yadav}, R.~K.~S., \& {Durgapal}, A.~K. 2017, \na, 52, 55,
  \dodoi{10.1016/j.newast.2016.10.009}

\bibitem[{Bisht {et~al.}(2021)Bisht, Zhu, Yadav, Ganesh, Rangwal, Durgapal,
  Sariya, \& Jiang}]{DB21}
Bisht, D., Zhu, Q., Yadav, R., {et~al.} 2021, \mnras, 503, 5929B,
  \dodoi{10.1093/mnras/stab691}

\bibitem[{{Boch} {et~al.}(2012){Boch}, {Pineau}, \& {Derriere}}]{CDS}
{Boch}, T., {Pineau}, F., \& {Derriere}, S. 2012, in Astronomical Society of
  the Pacific Conference Series, Vol. 461, Astronomical Data Analysis Software
  and Systems XXI, ed. P.~{Ballester}, D.~{Egret}, \& N.~P.~F. {Lorente}, 291

\bibitem[{{Bohlin} {et~al.}(1978){Bohlin}, {Savage}, \& {Drake}}]{Bohlin}
{Bohlin}, R.~C., {Savage}, B.~D., \& {Drake}, J.~F. 1978, \apj, 224, 132,
  \dodoi{10.1086/156357}

\bibitem[{{Bradley} {et~al.}(2020){Bradley}, {Sip{\H{o}}cz}, {Robitaille},
  {Tollerud}, {Vin{\'\i}cius}, {Deil}, {Barbary}, {Wilson}, {Busko},
  {G{\"u}nther}, {Cara}, {Conseil}, {Bostroem}, {Droettboom}, {Bray}, {Andersen
  Bratholm}, {Lim}, {Barentsen}, {Craig}, {Pascual}, {Perren}, {Greco},
  {Donath}, {De Val-Borro}, {Kerzendorf}, {Bach}, {Weaver}, {D'Eugenio},
  {Souchereau}, \& {Ferreira}}]{photutils}
{Bradley}, L., {Sip{\H{o}}cz}, B., {Robitaille}, T., {et~al.} 2020,
  {astropy/photutils: 1.0.0}, 1.0.0,  Zenodo, \dodoi{10.5281/zenodo.4044744}

\bibitem[{{Buckner} \& {Froebrich}(2013)}]{buckner2013properties}
{Buckner}, A. S.~M., \& {Froebrich}, D. 2013, \mnras, 436, 1465,
  \dodoi{10.1093/mnras/stt1665}

\bibitem[{{Cantat-Gaudin} {et~al.}(2018){Cantat-Gaudin}, {Jordi}, {Vallenari},
  {Bragaglia}, {Balaguer-N{\'u}{\~n}ez}, {Soubiran}, {Bossini}, {Moitinho},
  {Castro-Ginard}, {Krone-Martins}, {Casamiquela}, {Sordo}, \&
  {Carrera}}]{gaiaDR2OC}
{Cantat-Gaudin}, T., {Jordi}, C., {Vallenari}, A., {et~al.} 2018, \aap, 618,
  A93, \dodoi{10.1051/0004-6361/201833476}

\bibitem[{{Castro-Ginard} {et~al.}(2021){Castro-Ginard}, {McMillan}, {Luri},
  {Jordi}, {Romero-G{\'o}mez}, {Cantat-Gaudin}, {Casamiquela}, {Tarricq},
  {Soubiran}, \& {Anders}}]{spiral1}
{Castro-Ginard}, A., {McMillan}, P.~J., {Luri}, X., {et~al.} 2021, \aap, 652,
  A162, \dodoi{10.1051/0004-6361/202039751}

\bibitem[{{Chambers} {et~al.}(2016){Chambers}, {Magnier}, {Metcalfe},
  {Flewelling}, {Huber}, {Waters}, {Denneau}, {Draper}, {Farrow}, {Finkbeiner},
  {Holmberg}, {Koppenhoefer}, {Price}, {Rest}, {Saglia}, {Schlafly}, {Smartt},
  {Sweeney}, {Wainscoat}, {Burgett}, {Chastel}, {Grav}, {Heasley}, {Hodapp},
  {Jedicke}, {Kaiser}, {Kudritzki}, {Luppino}, {Lupton}, {Monet}, {Morgan},
  {Onaka}, {Shiao}, {Stubbs}, {Tonry}, {White}, {Ba{\~n}ados}, {Bell},
  {Bender}, {Bernard}, {Boegner}, {Boffi}, {Botticella}, {Calamida},
  {Casertano}, {Chen}, {Chen}, {Cole}, {Deacon}, {Frenk}, {Fitzsimmons},
  {Gezari}, {Gibbs}, {Goessl}, {Goggia}, {Gourgue}, {Goldman}, {Grant},
  {Grebel}, {Hambly}, {Hasinger}, {Heavens}, {Heckman}, {Henderson}, {Henning},
  {Holman}, {Hopp}, {Ip}, {Isani}, {Jackson}, {Keyes}, {Koekemoer}, {Kotak},
  {Le}, {Liska}, {Long}, {Lucey}, {Liu}, {Martin}, {Masci}, {McLean}, {Mindel},
  {Misra}, {Morganson}, {Murphy}, {Obaika}, {Narayan}, {Nieto-Santisteban},
  {Norberg}, {Peacock}, {Pier}, {Postman}, {Primak}, {Rae}, {Rai}, {Riess},
  {Riffeser}, {Rix}, {R{\"o}ser}, {Russel}, {Rutz}, {Schilbach}, {Schultz},
  {Scolnic}, {Strolger}, {Szalay}, {Seitz}, {Small}, {Smith}, {Soderblom},
  {Taylor}, {Thomson}, {Taylor}, {Thakar}, {Thiel}, {Thilker}, {Unger},
  {Urata}, {Valenti}, {Wagner}, {Walder}, {Walter}, {Watters}, {Werner},
  {Wood-Vasey}, \& {Wyse}}]{panstarrs1}
{Chambers}, K.~C., {Magnier}, E.~A., {Metcalfe}, N., {et~al.} 2016, arXiv
  e-prints, arXiv:1612.05560.
\newblock \doarXiv{1612.05560}

\bibitem[{{Clayton} {et~al.}(1995){Clayton}, {Wolff}, {Allen}, \&
  {Lupie}}]{Clayton1995}
{Clayton}, G.~C., {Wolff}, M.~J., {Allen}, R.~G., \& {Lupie}, O.~L. 1995, \apj,
  445, 947, \dodoi{10.1086/175754}

\bibitem[{{Clayton} {et~al.}(2003){Clayton}, {Wolff}, {Sofia}, {Gordon}, \&
  {Misselt}}]{clayton2003size}
{Clayton}, G.~C., {Wolff}, M.~J., {Sofia}, U.~J., {Gordon}, K.~D., \&
  {Misselt}, K.~A. 2003, \apj, 588, 871, \dodoi{10.1086/374316}

\bibitem[{{Clemens} {et~al.}(2020){Clemens}, {Cashman}, {Cerny}, {El-Batal},
  {Jameson}, {Marchwinski}, {Montgomery}, {Pavel}, {Pinnick}, \&
  {Taylor}}]{GPIPS}
{Clemens}, D.~P., {Cashman}, L.~R., {Cerny}, C., {et~al.} 2020, \apjs, 249, 23,
  \dodoi{10.3847/1538-4365/ab9f30}

\bibitem[{{Cutri} {et~al.}(2003){Cutri}, {Skrutskie}, {van Dyk}, {Beichman},
  {Carpenter}, {Chester}, {Cambresy}, {Evans}, {Fowler}, {Gizis}, {Howard},
  {Huchra}, {Jarrett}, {Kopan}, {Kirkpatrick}, {Light}, {Marsh}, {McCallon},
  {Schneider}, {Stiening}, {Sykes}, {Weinberg}, {Wheaton}, {Wheelock}, \&
  {Zacarias}}]{2MASS}
{Cutri}, R.~M., {Skrutskie}, M.~F., {van Dyk}, S., {et~al.} 2003, {2MASS All
  Sky Catalog of point sources.}

\bibitem[{{Dame} {et~al.}(2001){Dame}, {Hartmann}, \& {Thaddeus}}]{12CO}
{Dame}, T.~M., {Hartmann}, D., \& {Thaddeus}, P. 2001, \apj, 547, 792,
  \dodoi{10.1086/318388}

\bibitem[{{Dias} {et~al.}(2002){Dias}, {Alessi}, {Moitinho}, \&
  {L{\'e}pine}}]{Dias2002}
{Dias}, W.~S., {Alessi}, B.~S., {Moitinho}, A., \& {L{\'e}pine}, J.~R.~D. 2002,
  \aap, 389, 871, \dodoi{10.1051/0004-6361:20020668}

\bibitem[{{Dutra} \& {Bica}(2002)}]{Dutra2002}
{Dutra}, C.~M., \& {Bica}, E. 2002, \aap, 383, 631,
  \dodoi{10.1051/0004-6361:20011761}

\bibitem[{{Eswaraiah} {et~al.}(2012){Eswaraiah}, {Pandey}, {Maheswar}, {Chen},
  {Ojha}, \& {Chandola}}]{Bk59}
{Eswaraiah}, C., {Pandey}, A.~K., {Maheswar}, G., {et~al.} 2012, \mnras, 419,
  2587, \dodoi{10.1111/j.1365-2966.2011.19908.x}

\bibitem[{{Eswaraiah} {et~al.}(2011){Eswaraiah}, {Pandey}, {Maheswar}, {Medhi},
  {Pandey}, {Ojha}, \& {Chen}}]{NGC1893}
---. 2011, \mnras, 411, 1418, \dodoi{10.1111/j.1365-2966.2010.17780.x}

\bibitem[{{Fabricius} {et~al.}(2021){Fabricius}, {Luri}, {Arenou}, {Babusiaux},
  {Helmi}, {Muraveva}, {Reyl{\'e}}, {Spoto}, {Vallenari}, {Antoja}, {Balbinot},
  {Barache}, {Bauchet}, {Bragaglia}, {Busonero}, {Cantat-Gaudin}, {Carrasco},
  {Diakit{\'e}}, {Fabrizio}, {Figueras}, {Garcia-Gutierrez}, {Garofalo},
  {Jordi}, {Kervella}, {Khanna}, {Leclerc}, {Licata}, {Lambert}, {Marrese},
  {Masip}, {Ramos}, {Robichon}, {Robin}, {Romero-G{\'o}mez}, {Rubele}, \&
  {Weiler}}]{GaiaEDR3Cat}
{Fabricius}, C., {Luri}, X., {Arenou}, F., {et~al.} 2021, \aap, 649, A5,
  \dodoi{10.1051/0004-6361/202039834}

\bibitem[{{Feinstein} {et~al.}(2000){Feinstein}, {Baume}, {Vazquez}, {Niemela},
  \& {Cerruti}}]{Trumpler27}
{Feinstein}, C., {Baume}, G., {Vazquez}, R., {Niemela}, V., \& {Cerruti}, M.~A.
  2000, \aj, 120, 1906, \dodoi{10.1086/301562}

\bibitem[{{Gaia Collaboration} {et~al.}(2021){Gaia Collaboration}, {Brown},
  {Vallenari}, {Prusti}, {de Bruijne}, {Babusiaux}, {Biermann}, {Creevey},
  {Evans}, {Eyer}, {Hutton}, {Jansen}, {Jordi}, {Klioner}, {Lammers},
  {Lindegren}, {Luri}, {Mignard}, {Panem}, {Pourbaix}, {Randich}, {Sartoretti},
  {Soubiran}, {Walton}, {Arenou}, {Bailer-Jones}, {Bastian}, {Cropper},
  {Drimmel}, {Katz}, {Lattanzi}, {van Leeuwen}, {Bakker}, {Cacciari},
  {Casta{\~n}eda}, {De Angeli}, {Ducourant}, {Fabricius}, {Fouesneau},
  {Fr{\'e}mat}, {Guerra}, {Guerrier}, {Guiraud}, {Jean-Antoine Piccolo},
  {Masana}, {Messineo}, {Mowlavi}, {Nicolas}, {Nienartowicz}, {Pailler},
  {Panuzzo}, {Riclet}, {Roux}, {Seabroke}, {Sordo}, {Tanga}, {Th{\'e}venin},
  {Gracia-Abril}, {Portell}, {Teyssier}, {Altmann}, {Andrae}, {Bellas-Velidis},
  {Benson}, {Berthier}, {Blomme}, {Brugaletta}, {Burgess}, {Busso}, {Carry},
  {Cellino}, {Cheek}, {Clementini}, {Damerdji}, {Davidson}, {Delchambre},
  {Dell'Oro}, {Fern{\'a}ndez-Hern{\'a}ndez}, {Galluccio}, {Garc{\'\i}a-Lario},
  {Garcia-Reinaldos}, {Gonz{\'a}lez-N{\'u}{\~n}ez}, {Gosset}, {Haigron},
  {Halbwachs}, {Hambly}, {Harrison}, {Hatzidimitriou}, {Heiter},
  {Hern{\'a}ndez}, {Hestroffer}, {Hodgkin}, {Holl}, {Jan{\ss}en}, {Jevardat de
  Fombelle}, {Jordan}, {Krone-Martins}, {Lanzafame}, {L{\"o}ffler}, {Lorca},
  {Manteiga}, {Marchal}, {Marrese}, {Moitinho}, {Mora}, {Muinonen}, {Osborne},
  {Pancino}, {Pauwels}, {Petit}, {Recio-Blanco}, {Richards}, {Riello},
  {Rimoldini}, {Robin}, {Roegiers}, {Rybizki}, {Sarro}, {Siopis}, {Smith},
  {Sozzetti}, {Ulla}, {Utrilla}, {van Leeuwen}, {van Reeven}, {Abbas}, {Abreu
  Aramburu}, {Accart}, {Aerts}, {Aguado}, {Ajaj}, {Altavilla}, {{\'A}lvarez},
  {{\'A}lvarez Cid-Fuentes}, {Alves}, {Anderson}, {Anglada Varela}, {Antoja},
  {Audard}, {Baines}, {Baker}, {Balaguer-N{\'u}{\~n}ez}, {Balbinot}, {Balog},
  {Barache}, {Barbato}, {Barros}, {Barstow}, {Bartolom{\'e}}, {Bassilana},
  {Bauchet}, {Baudesson-Stella}, {Becciani}, {Bellazzini}, {Bernet}, {Bertone},
  {Bianchi}, {Blanco-Cuaresma}, {Boch}, {Bombrun}, {Bossini}, {Bouquillon},
  {Bragaglia}, {Bramante}, {Breedt}, {Bressan}, {Brouillet}, {Bucciarelli},
  {Burlacu}, {Busonero}, {Butkevich}, {Buzzi}, {Caffau}, {Cancelliere},
  {C{\'a}novas}, {Cantat-Gaudin}, {Carballo}, {Carlucci}, {Carnerero},
  {Carrasco}, {Casamiquela}, {Castellani}, {Castro-Ginard}, {Castro Sampol},
  {Chaoul}, {Charlot}, {Chemin}, {Chiavassa}, {Cioni}, {Comoretto}, {Cooper},
  {Cornez}, {Cowell}, {Crifo}, {Crosta}, {Crowley}, {Dafonte}, {Dapergolas},
  {David}, {David}, {de Laverny}, {De Luise}, {De March}, {De Ridder}, {de
  Souza}, {de Teodoro}, {de Torres}, {del Peloso}, {del Pozo}, {Delbo},
  {Delgado}, {Delgado}, {Delisle}, {Di Matteo}, {Diakite}, {Diener},
  {Distefano}, {Dolding}, {Eappachen}, {Edvardsson}, {Enke}, {Esquej}, {Fabre},
  {Fabrizio}, {Faigler}, {Fedorets}, {Fernique}, {Fienga}, {Figueras},
  {Fouron}, {Fragkoudi}, {Fraile}, {Franke}, {Gai}, {Garabato},
  {Garcia-Gutierrez}, {Garc{\'\i}a-Torres}, {Garofalo}, {Gavras}, {Gerlach},
  {Geyer}, {Giacobbe}, {Gilmore}, {Girona}, {Giuffrida}, {Gomel}, {Gomez},
  {Gonzalez-Santamaria}, {Gonz{\'a}lez-Vidal}, {Granvik},
  {Guti{\'e}rrez-S{\'a}nchez}, {Guy}, {Hauser}, {Haywood}, {Helmi}, {Hidalgo},
  {Hilger}, {H{\l}adczuk}, {Hobbs}, {Holland}, {Huckle}, {Jasniewicz},
  {Jonker}, {Juaristi Campillo}, {Julbe}, {Karbevska}, {Kervella}, {Khanna},
  {Kochoska}, {Kontizas}, {Kordopatis}, {Korn}, {Kostrzewa-Rutkowska},
  {Kruszy{\'n}ska}, {Lambert}, {Lanza}, {Lasne}, {Le Campion}, {Le Fustec},
  {Lebreton}, {Lebzelter}, {Leccia}, {Leclerc}, {Lecoeur-Taibi}, {Liao},
  {Licata}, {Lindstr{\o}m}, {Lister}, {Livanou}, {Lobel}, {Madrero Pardo},
  {Managau}, {Mann}, {Marchant}, {Marconi}, {Marcos Santos}, {Marinoni},
  {Marocco}, {Marshall}, {Martin Polo}, {Mart{\'\i}n-Fleitas}, {Masip},
  {Massari}, {Mastrobuono-Battisti}, {Mazeh}, {McMillan}, {Messina},
  {Michalik}, {Millar}, {Mints}, {Molina}, {Molinaro}, {Moln{\'a}r},
  {Montegriffo}, {Mor}, {Morbidelli}, {Morel}, {Morris}, {Mulone}, {Munoz},
  {Muraveva}, {Murphy}, {Musella}, {Noval}, {Ord{\'e}novic}, {Orr{\`u}},
  {Osinde}, {Pagani}, {Pagano}, {Palaversa}, {Palicio}, {Panahi}, {Pawlak},
  {Pe{\~n}alosa Esteller}, {Penttil{\"a}}, {Piersimoni}, {Pineau}, {Plachy},
  {Plum}, {Poggio}, {Poretti}, {Poujoulet}, {Pr{\v{s}}a}, {Pulone}, {Racero},
  {Ragaini}, {Rainer}, {Raiteri}, {Rambaux}, {Ramos}, {Ramos-Lerate}, {Re
  Fiorentin}, {Regibo}, {Reyl{\'e}}, {Ripepi}, {Riva}, {Rixon}, {Robichon},
  {Robin}, {Roelens}, {Rohrbasser}, {Romero-G{\'o}mez}, {Rowell}, {Royer},
  {Rybicki}, {Sadowski}, {Sagrist{\`a} Sell{\'e}s}, {Sahlmann}, {Salgado},
  {Salguero}, {Samaras}, {Sanchez Gimenez}, {Sanna}, {Santove{\~n}a},
  {Sarasso}, {Schultheis}, {Sciacca}, {Segol}, {Segovia}, {S{\'e}gransan},
  {Semeux}, {Shahaf}, {Siddiqui}, {Siebert}, {Siltala}, {Slezak}, {Smart},
  {Solano}, {Solitro}, {Souami}, {Souchay}, {Spagna}, {Spoto}, {Steele},
  {Steidelm{\"u}ller}, {Stephenson}, {S{\"u}veges}, {Szabados}, {Szegedi-Elek},
  {Taris}, {Tauran}, {Taylor}, {Teixeira}, {Thuillot}, {Tonello}, {Torra},
  {Torra}, {Turon}, {Unger}, {Vaillant}, {van Dillen}, {Vanel}, {Vecchiato},
  {Viala}, {Vicente}, {Voutsinas}, {Weiler}, {Wevers}, {Wyrzykowski}, {Yoldas},
  {Yvard}, {Zhao}, {Zorec}, {Zucker}, {Zurbach}, \& {Zwitter}}]{EDR3}
{Gaia Collaboration}, {Brown}, A.~G.~A., {Vallenari}, A., {et~al.} 2021, \aap,
  649, A1, \dodoi{10.1051/0004-6361/202039657}

\bibitem[{{Ganesh} {et~al.}(2020){Ganesh}, {Rai}, {Aravind}, {Singh},
  {Prajapati}, {Mishra}, {Kasarla}, {Sarkar}, {Patwal}, {Uppal}, {Chandra},
  {Mathur}, {Shah}, {Baliyan}, \& {Joshi}}]{EMPOL}
{Ganesh}, S., {Rai}, A., {Aravind}, K., {et~al.} 2020, in Society of
  Photo-Optical Instrumentation Engineers (SPIE) Conference Series, Vol. 11447,
  Society of Photo-Optical Instrumentation Engineers (SPIE) Conference Series,
  114479E, \dodoi{10.1117/12.2560949}

\bibitem[{Girard {et~al.}(1989)Girard, Grundy, Lopez, \& van Altena}]{Girard89}
Girard, T.~M., Grundy, W.~M., Lopez, C.~E., \& van Altena, W.~F. 1989, \aj, 98,
  227G, \dodoi{10.1086/115139}

\bibitem[{{Gonz{\'a}lez-Fern{\'a}ndez}
  {et~al.}(2018){Gonz{\'a}lez-Fern{\'a}ndez}, {Hodgkin}, {Irwin},
  {Gonz{\'a}lez-Solares}, {Koposov}, {Lewis}, {Emerson}, {Hewett},
  {Yolda{\c{s}}}, \& {Riello}}]{2MASSVVV}
{Gonz{\'a}lez-Fern{\'a}ndez}, C., {Hodgkin}, S.~T., {Irwin}, M.~J., {et~al.}
  2018, \mnras, 474, 5459, \dodoi{10.1093/mnras/stx3073}

\bibitem[{{Goodman} {et~al.}(1990){Goodman}, {Bastien}, {Myers}, \&
  {Menard}}]{goodman1990optical}
{Goodman}, A.~A., {Bastien}, P., {Myers}, P.~C., \& {Menard}, F. 1990, \apj,
  359, 363, \dodoi{10.1086/169070}

\bibitem[{{Green}(2018)}]{dustPython}
{Green}, G.~M. 2018, The Journal of Open Source Software, 3, 695,
  \dodoi{10.21105/joss.00695}

\bibitem[{{Green} {et~al.}(2019){Green}, {Schlafly}, {Zucker}, {Speagle}, \&
  {Finkbeiner}}]{Green2019}
{Green}, G.~M., {Schlafly}, E., {Zucker}, C., {Speagle}, J.~S., \&
  {Finkbeiner}, D. 2019, \apj, 887, 93, \dodoi{10.3847/1538-4357/ab5362}

\bibitem[{{Hao} {et~al.}(2021){Hao}, {Xu}, {Hou}, {Bian}, {Li}, {Wu}, {He},
  {Li}, \& {Liu}}]{spiral2}
{Hao}, C.~J., {Xu}, Y., {Hou}, L.~G., {et~al.} 2021, \aap, 652, A102,
  \dodoi{10.1051/0004-6361/202140608}

\bibitem[{{Heiles}(2000)}]{Heiles2000}
{Heiles}, C. 2000, \aj, 119, 923, \dodoi{10.1086/301236}

\bibitem[{{HI4PI Collaboration} {et~al.}(2016){HI4PI Collaboration}, {Ben
  Bekhti}, {Fl{\"o}er}, {Keller}, {Kerp}, {Lenz}, {Winkel}, {Bailin},
  {Calabretta}, {Dedes}, {Ford}, {Gibson}, {Haud}, {Janowiecki}, {Kalberla},
  {Lockman}, {McClure-Griffiths}, {Murphy}, {Nakanishi}, {Pisano}, \&
  {Staveley-Smith}}]{HI4PI}
{HI4PI Collaboration}, {Ben Bekhti}, N., {Fl{\"o}er}, L., {et~al.} 2016, \aap,
  594, A116, \dodoi{10.1051/0004-6361/201629178}

\bibitem[{{Hiltner}(1956)}]{Hiltner1956}
{Hiltner}, W.~A. 1956, \apjs, 2, 389, \dodoi{10.1086/190029}

\bibitem[{{Joshi} {et~al.}(2016){Joshi}, {Dambis}, {Pandey}, \&
  {Joshi}}]{JoshiGC}
{Joshi}, Y.~C., {Dambis}, A.~K., {Pandey}, A.~K., \& {Joshi}, S. 2016, \aap,
  593, A116, \dodoi{10.1051/0004-6361/201628944}

\bibitem[{{Kharchenko} {et~al.}(2013){Kharchenko}, {Piskunov}, {Schilbach},
  {R{\"o}ser}, \& {Scholz}}]{kharchenko2013}
{Kharchenko}, N.~V., {Piskunov}, A.~E., {Schilbach}, E., {R{\"o}ser}, S., \&
  {Scholz}, R.~D. 2013, \aap, 558, A53, \dodoi{10.1051/0004-6361/201322302}

\bibitem[{{Kharchenko} {et~al.}(2016){Kharchenko}, {Piskunov}, {Schilbach},
  {R{\"o}ser}, \& {Scholz}}]{Kharchenko2016}
---. 2016, \aap, 585, A101, \dodoi{10.1051/0004-6361/201527292}

\bibitem[{{Kim} \& {Martin}(1994)}]{kim1994size}
{Kim}, S.-H., \& {Martin}, P.~G. 1994, \apj, 431, 783, \dodoi{10.1086/174529}

\bibitem[{{Kim} \& {Martin}(1995)}]{kim1995size}
---. 1995, \apj, 444, 293, \dodoi{10.1086/175604}

\bibitem[{{Lallement} {et~al.}(2019){Lallement}, {Babusiaux}, {Vergely},
  {Katz}, {Arenou}, {Valette}, {Hottier}, \& {Capitanio}}]{Lallement2019}
{Lallement}, R., {Babusiaux}, C., {Vergely}, J.~L., {et~al.} 2019, \aap, 625,
  A135, \dodoi{10.1051/0004-6361/201834695}

\bibitem[{{Lynds}(1962)}]{Lynds1962}
{Lynds}, B.~T. 1962, \apjs, 7, 1, \dodoi{10.1086/190072}

\bibitem[{{Martin} {et~al.}(1999){Martin}, {Clayton}, \& {Wolff}}]{martin}
{Martin}, P.~G., {Clayton}, G.~C., \& {Wolff}, M.~J. 1999, \apj, 510, 905,
  \dodoi{10.1086/306613}

\bibitem[{{Mart{\'\i}nez} {et~al.}(2004){Mart{\'\i}nez}, {Vergne}, \&
  {Feinstein}}]{Hogg22}
{Mart{\'\i}nez}, R., {Vergne}, M.~M., \& {Feinstein}, C. 2004, \aap, 419, 965,
  \dodoi{10.1051/0004-6361:20035635}

\bibitem[{{Marton} {et~al.}(2019){Marton}, {{\'A}brah{\'a}m}, {Szegedi-Elek},
  {Varga}, {Kun}, {K{\'o}sp{\'a}l}, {Varga-Vereb{\'e}lyi}, {Hodgkin},
  {Szabados}, {Beck}, \& {Kiss}}]{YSOs}
{Marton}, G., {{\'A}brah{\'a}m}, P., {Szegedi-Elek}, E., {et~al.} 2019, \mnras,
  487, 2522, \dodoi{10.1093/mnras/stz1301}

\bibitem[{{Mathewson} \& {Ford}(1971)}]{mathewson1970polarization}
{Mathewson}, D.~S., \& {Ford}, V.~L. 1971, \mnras, 153, 525,
  \dodoi{10.1093/mnras/153.4.525}

\bibitem[{{Mathis}(1986)}]{mathis1986}
{Mathis}, J.~S. 1986, \apj, 308, 281, \dodoi{10.1086/164499}

\bibitem[{{Medhi} {et~al.}(2007){Medhi}, {Maheswar}, {Brijesh}, {Pandey},
  {Kumar}, \& {Sagar}}]{IC1805}
{Medhi}, B.~J., {Maheswar}, G., {Brijesh}, K., {et~al.} 2007, \mnras, 378, 881,
  \dodoi{10.1111/j.1365-2966.2007.11767.x}

\bibitem[{{Medhi} {et~al.}(2008){Medhi}, {Maheswar}, {Pandey}, {Kumar}, \&
  {Sagar}}]{NGC654}
{Medhi}, B.~J., {Maheswar}, G., {Pandey}, J.~C., {Kumar}, T.~S., \& {Sagar}, R.
  2008, \mnras, 388, 105, \dodoi{10.1111/j.1365-2966.2008.13405.x}

\bibitem[{{Medhi} {et~al.}(2010){Medhi}, {Maheswar G.}, {Pandey}, {Tamura}, \&
  {Sagar}}]{NGC6823}
{Medhi}, B.~J., {Maheswar G.}, {Pandey}, J.~C., {Tamura}, M., \& {Sagar}, R.
  2010, \mnras, 403, 1577, \dodoi{10.1111/j.1365-2966.2010.16221.x}

\bibitem[{{Medhi} \& {Tamura}(2013)}]{membership}
{Medhi}, B.~J., \& {Tamura}, M. 2013, \mnras, 430, 1334,
  \dodoi{10.1093/mnras/sts714}

\bibitem[{{Mink}(2011)}]{WCSTOOL}
{Mink}, J. 2011, {WCSTools: Image Astrometry Toolkit}.
\newblock \doeprint{1109.015}

\bibitem[{{NASA/IPAC Infrared Science Archive}(2020)}]{WISEW4}
{NASA/IPAC Infrared Science Archive}. 2020, WISE All-Sky 4-band Atlas Coadded
  Images,  IPAC, \dodoi{10.26131/IRSA151}

\bibitem[{Pancharatnam(1955)}]{panch}
Pancharatnam, S. 1955, Proceedings of the Indian Academy of Sciences - Section
  A, 41, 137, \dodoi{10.1007/bf03047098}

\bibitem[{{Papoular}(2018)}]{NewSerkowski}
{Papoular}, R. 2018, \mnras, 479, 1685, \dodoi{10.1093/mnras/sty1530}

\bibitem[{{Reid} {et~al.}(2009){Reid}, {Menten}, {Zheng}, {Brunthaler},
  {Moscadelli}, {Xu}, {Zhang}, {Sato}, {Honma}, {Hirota}, {Hachisuka}, {Choi},
  {Moellenbrock}, \& {Bartkiewicz}}]{Reid2009KinematicDistance}
{Reid}, M.~J., {Menten}, K.~M., {Zheng}, X.~W., {et~al.} 2009, \apj, 700, 137,
  \dodoi{10.1088/0004-637X/700/1/137}

\bibitem[{{Reid} {et~al.}(2014){Reid}, {Menten}, {Brunthaler}, {Zheng}, {Dame},
  {Xu}, {Wu}, {Zhang}, {Sanna}, {Sato}, {Hachisuka}, {Choi}, {Immer},
  {Moscadelli}, {Rygl}, \& {Bartkiewicz}}]{Reid2014parameters}
{Reid}, M.~J., {Menten}, K.~M., {Brunthaler}, A., {et~al.} 2014, \apj, 783,
  130, \dodoi{10.1088/0004-637X/783/2/130}

\bibitem[{Sariya {et~al.}(2021)Sariya, Jiang, Bisht, Yadav, \& Rangwal}]{DPS21}
Sariya, D., Jiang, I.-G., Bisht, D., Yadav, R., \& Rangwal, G. 2021, \aj, 161,
  102S, \dodoi{10.3847/1538-3881/abd31f}

\bibitem[{{Schmidt} {et~al.}(1992){Schmidt}, {Elston}, \& {Lupie}}]{HD25443}
{Schmidt}, G.~D., {Elston}, R., \& {Lupie}, O.~L. 1992, \aj, 104, 1563,
  \dodoi{10.1086/116341}

\bibitem[{{Seaton}(1979)}]{Seaton}
{Seaton}, M.~J. 1979, \mnras, 187, 73, \dodoi{10.1093/mnras/187.1.73P}

\bibitem[{{Serkowski} {et~al.}(1975){Serkowski}, {Mathewson}, \&
  {Ford}}]{Serkowski1975}
{Serkowski}, K., {Mathewson}, D.~S., \& {Ford}, V.~L. 1975, \apj, 196, 261,
  \dodoi{10.1086/153410}

\bibitem[{Shao \& Zhao(1996)}]{shao96}
Shao, Z., \& Zhao, J. 1996, Acta Astron. Sinica, 37, 377S

\bibitem[{{Sharma} {et~al.}(2020){Sharma}, {Ghosh}, {Ojha}, {Pandey}, {Sinha},
  {Pandey}, {Ghosh}, {Panwar}, \& {Pandey}}]{CZ3sharma2020}
{Sharma}, S., {Ghosh}, A., {Ojha}, D.~K., {et~al.} 2020, \mnras, 498, 2309,
  \dodoi{10.1093/mnras/staa2412}

\bibitem[{{Singh} \& {Pandey}(2020)}]{NGC1817}
{Singh}, S., \& {Pandey}, J.~C. 2020, \aj, 160, 256,
  \dodoi{10.3847/1538-3881/abba29}

\bibitem[{{Singh} {et~al.}(2020){Singh}, {Pandey}, {Yadav}, \&
  {Medhi}}]{Alessi1}
{Singh}, S., {Pandey}, J.~C., {Yadav}, R.~K.~S., \& {Medhi}, B.~J. 2020, \aj,
  159, 99, \dodoi{10.3847/1538-3881/ab6608}

\bibitem[{{Topasna} {et~al.}(2017){Topasna}, {Daman}, \& {Kaltcheva}}]{NGC457}
{Topasna}, G.~A., {Daman}, E.~A., \& {Kaltcheva}, N.~T. 2017, \pasp, 129,
  104201, \dodoi{10.1088/1538-3873/aa7f6c}

\bibitem[{{Topasna} {et~al.}(2018){Topasna}, {Kaltcheva}, \&
  {Paunzen}}]{NGC1502}
{Topasna}, G.~A., {Kaltcheva}, N.~T., \& {Paunzen}, E. 2018, \aap, 615, A166,
  \dodoi{10.1051/0004-6361/201731903}

\bibitem[{{Vergne} {et~al.}(2007){Vergne}, {Feinstein}, \&
  {Mart{\'\i}nez}}]{NGC5749}
{Vergne}, M.~M., {Feinstein}, C., \& {Mart{\'\i}nez}, R. 2007, \aap, 462, 621,
  \dodoi{10.1051/0004-6361:20042124}

\bibitem[{{Voshchinnikov} \& {Das}(2008)}]{efficiency}
{Voshchinnikov}, N.~V., \& {Das}, H.~K. 2008, \jqsrt, 109, 1527,
  \dodoi{10.1016/j.jqsrt.2008.01.003}

\bibitem[{Whittet(2003)}]{Whittet2003}
Whittet, D. 2003, Dust in the galactic environment

\bibitem[{{Whittet}(1992)}]{whittet1992dust}
{Whittet}, D.~C.~B. 1992, {Dust in the galactic environment}

\bibitem[{{Wilking} {et~al.}(1980){Wilking}, {Lebofsky}, {Martin}, {Rieke}, \&
  {Kemp}}]{Wilking1980}
{Wilking}, B.~A., {Lebofsky}, M.~J., {Martin}, P.~G., {Rieke}, G.~H., \&
  {Kemp}, J.~C. 1980, \apj, 235, 905, \dodoi{10.1086/157694}

\bibitem[{{Wilking} {et~al.}(1982){Wilking}, {Lebofsky}, \&
  {Rieke}}]{Wilking1982}
{Wilking}, B.~A., {Lebofsky}, M.~J., \& {Rieke}, G.~H. 1982, \aj, 87, 695,
  \dodoi{10.1086/113147}

\bibitem[{{Xu} {et~al.}(2006){Xu}, {Reid}, {Zheng}, \& {Menten}}]{perseusDist}
{Xu}, Y., {Reid}, M.~J., {Zheng}, X.~W., \& {Menten}, K.~M. 2006, Science, 311,
  54, \dodoi{10.1126/science.1120914}

\bibitem[{{Zacharias} {et~al.}(2013){Zacharias}, {Finch}, {Girard}, {Henden},
  {Bartlett}, {Monet}, \& {Zacharias}}]{UCACcat}
{Zacharias}, N., {Finch}, C.~T., {Girard}, T.~M., {et~al.} 2013, \aj, 145, 44,
  \dodoi{10.1088/0004-6256/145/2/44}

\bibitem[{{Zucker} {et~al.}(2020){Zucker}, {Speagle}, {Schlafly}, {Green},
  {Finkbeiner}, {Goodman}, \& {Alves}}]{darkcloudDist2020}
{Zucker}, C., {Speagle}, J.~S., {Schlafly}, E.~F., {et~al.} 2020, \aap, 633,
  A51, \dodoi{10.1051/0004-6361/201936145}

\end{thebibliography}
\bibliographystyle{aasjournal}
\end{document}